# On Approximating Frequency Moments of Data Streams with Skewed Projections

Ping Li  (pingli@cornell.edu)     Department of Statistical Science

Faculty of Computing and Information Science

Cornell University, Ithaca, NY 14853

November 19, 2007 (revised October 26, 2018)

## Abstract

We propose *skewed stable random projections* for approximating the $\alpha$th frequency moments of dynamic data streams ($0 < \alpha \leq 2$). We show the sample complexity (number of projections) $k = G\frac{1}{\epsilon^2}\log\left(\frac{2}{\delta}\right)$, where $G \to \frac{\epsilon^2}{\log(1+\epsilon)} = O(\epsilon)$ as $\alpha \to 1$, i.e., $\alpha = 1 \pm \Delta$ with $\Delta \to 0$. Previous results based on *symmetric stable random projections*[12, 16] required $G =$ non-zero constant $+ O(\epsilon)$, even when $\Delta = 0$. The case $\Delta \to 0$ is practically important. For example, $\Delta$ might be the "decay rate" or "interest rate," which is usually small; and hence one might view *skewed stable random projections* as a "generalized counter" for estimating the total value in the future, taking in account of the effect of decaying or interest accruement.

We consider the popular *Turnstile* data stream model. The input data stream $a_t = (i, I_t)$ arriving sequentially describes the underlying signal $A$, meaning $A_t[i] = A_{t-1}[i] + I_t$, $i \in [1, D]$. We allow the increment $I_t$ to be either positive (i.e., insertion) or negative (i.e., deletion). By definition, the $\alpha$th frequency moment $F_{(\alpha)} = \sum_{i=1}^{D}|A_t(i)|^\alpha$. Our method only requires that, at the time $t$ for the evaluation, $A_t(i) \geq 0$, which is only a minor restriction for natural data streams encountered in practice.

More specifically, compared with previous studies[11, 12, 16], our contributions are two-fold.

1. **Our proposal of *skewed stable random projections* for data stream computations**
   In FOCS'00[11], Indyk proposed *(symmetric) stable random projections* for approximating the $\alpha$th frequency moment of data streams, where $0 < \alpha \leq 2$. Because practical data streams are often: (a) insertion only (i.e., the *cash register* model), or (b) always non-negative (i.e., the *strict Turnstile* model), or (c) ultimately non-negative at check points, using *symmetric stable random projections* is often not necessary. Consider at the time $t$, $A_t(i) \geq 0$ for all $i$. When $\alpha = 1$, we can compute $F_{(1)}$ essentially error-free using a counter. However, if one applies *symmetric stable random projections* and the *geometric mean* estimator in [16], the sample complexity requires $k = \left(\frac{\pi^2}{2} + O(\epsilon)\right)\frac{1}{\epsilon^2}\log\frac{2}{\delta}$. The situation becomes much more interesting when $\alpha = 1 \pm \Delta$ with small $\Delta$, because in this case the traditional counter can not be used but *symmetric stable random projections* will still require a large number of samples (projections).

   For the first time, we propose *skewed stable random projections*, which may be viewed as a "generalized counter" and works especially well when $\Delta$ is small, which is also practically very important.

2. **Our development of various statistical estimators for skewed stable distributions**
   Good statistical estimators are both theoretically important (e.g., for sample complexity bounds) and practically useful (e.g., for accurate estimates using fewer samples). The method of *skewed stable random projections* eventually boils down to a statistical estimation problem, which is less well-studied in statistics than for *symmetric stable random projections*. Thus, much of our work is based on the first principle.

   - To build the foundation for statistical estimation, we derive theoretical formulas for moments of skewed stable distributions and discover a useful property that a *fully skewed* stable distribution has infinite-order negative moments. We only recommend *fully skewed* projections.
   - We design a general estimator based on the *geometric mean* for skewed stable distributions and show that the estimation variance is minimized in *fully skewed* stable distributions. The asymptotic variance of the estimator is $\frac{(1-\alpha^2)\pi^2}{6}\frac{1}{k}F_{(\alpha)}^2$ (when $\alpha < 1$) and $\frac{(5-\alpha)(\alpha-1)\pi^2}{6}\frac{1}{k}F_{(\alpha)}^2$ (when $\alpha > 1$). Compared with [16], our work in a sense achieves an *infinite* improvement when $\alpha \to 1$, in terms of the asymptotic variances. We also provide explicit tail bounds and consequently establish that $k = G\frac{1}{\epsilon^2}\log\left(\frac{2}{\delta}\right)$, where $G = \frac{\epsilon^2}{\log(1+\epsilon) - 2\sqrt{\Delta\log(1+\epsilon)} + o(\sqrt{\Delta})}$ as $\alpha = 1 \pm \Delta \to 1$ (i.e., $\Delta \to 0$).
   - For $\alpha < 1$, the *harmonic mean* estimator is considerably more accurate. Unlike the *harmonic mean* estimator in [16] (which was useful only for very small $\alpha$), this estimator has infinite-order moments and hence exhibits nice tail behaviors for all $0 < \alpha < 1$. We provide the tail bounds explicitly.

- Maximum likelihood estimators (MLE) can be explicitly derived for $\alpha = 0+$, $\alpha = 0.5$ and $\alpha = 2$. We analyze the MLE for $\alpha = 0.5$, including the variances and explicit tail bounds.
- Finally, we also propose the *optimal power* estimator, which becomes the MLE when $\alpha = 0.5$, $0+$, or 2. Moreover, for $\alpha < 1$, all moments exist and exponential bounds can be established.

# 1 Introduction

The ubiquitous phenomenon of massive data streams[10, 7, 12, 2, 6, 19] imposes many challenges including *transmit*, *compute*, and *store*[19]. In fact, "Scaling Up for High Dimensional Data and High Speed Data Streams" is among the "ten challenging problems in data mining research."[1] This paper focuses on approximating frequency moments of streams, using a new method called *skewed stable random projections*, which considerably (or even "infinitely" in special cases) improves previous methods based on *symmetric stable random projections*[11, 12, 16].

Consider the popular *Turnstile* model [19]. The input data stream $a_t = (i, I_t)$ arriving sequentially describes the underlying signal $A$, meaning $A_t[i] = A_{t-1}[i] + I_t$, $i \in [1, D]$. The increment $I_t$ can be either positive (insertion) or negative (deletion). Restricting $I_t \geq 0$ results in the *cash register* model. Restricting $A_t[i] \geq 0$ at all times $t$ (but still allowing $I_t$ to be either positive or negative) results in the *strict Turnstile* model, which suffices for describing many (but not all) natural phenomena. For example[19], in a database, a record can only be deleted if it was previously inserted. Another example is the checking/savings account, which allows deposits and withdrawals but in generally does not allow overdraft.

Our proposed method of *skewed table random projections* is applicable when, at the time $t$ for the evaluation, $A_t[i] \geq 0$ for all $i$. This is much more flexible than the *strict Turnstile* model, which requires that $A_t[i] \geq 0$ for all $t$. In other words, our proposed method is applicable to data streams that are (a) insertion only (i.e., the *cash register* model), or (b) always non-negative (i.e., the *strict Turnstile* model), or (c) eventually non-negative at check points. We believe our model suffices for most natural data streams encountered in practice.

Pioneered by[1], there have been many studies on approximating the $\alpha$th frequency moment $F_{(\alpha)}$, defined as

$$F_{(\alpha)} = \sum_{i=1}^{D} (A_t[i])^\alpha.$$

[1] considered integer moments, $\alpha = 0, 1, 2$, as well as $\alpha > 2$. Soon after, [7, 11] provided improved algorithms for $0 < \alpha \leq 2$. [20, 3] proved the sample complexity lower bounds for $\alpha > 2$. [23] proved the optimal lower bounds for all frequency moments, except for $\alpha = 1$, because [23] considered non-negative data streams ($A_t[i] \geq 0$), for which one can compute $F_{(1)}$ essentially error-free with a counter[18, 8, 1]. [13] provided algorithms for $\alpha > 2$ to (essentially) achieve the lower bounds proved in [20, 3]. We should also mention that the fundamental complexity results [24, 25] were used in the proofs in [1, 20, 3, 23].

Our proposed method of *skewed stable random projections* is applicable when $0 < \alpha \leq 2$ and it works particularly well when $\alpha$ is only slightly smaller or larger than 1, i.e., $\alpha = 1 \pm \Delta$ and $\Delta$ is small. This can be practically very useful. For example, $\Delta$ may be interpreted as the "decay rate" or the "interest rate," which is usually small. In a sense, we can view *skewed stable random projections* as a "generalized counter" in that it can count the total values in the future taking into account the effect of decaying or interest accruement.

This is the first paper on *skewed stable random projections*, and hence we start with a brief introduction to skewed stable distributions.

## 1.1 Skewed Stable Distributions

A random variable $Z$ follows a $\beta$-skewed $\alpha$-stable distribution if the Fourier transform of its density is[26, 21]

$$\mathscr{F}_Z(t) = \mathrm{E} \exp\left(\sqrt{-1} Z t\right) = \exp\left(-F|t|^\alpha \left(1 - \sqrt{-1}\beta \mathrm{sgn}(t) \tan\left(\frac{\pi\alpha}{2}\right)\right)\right), \qquad \alpha \neq 1,$$

where $-1 \leq \beta \leq 1$ and $F > 0$ is the scale parameter. We denote $Z \sim S(\alpha, \beta, F)$.

---
[1]http://www.cs.uvm.edu/~icdm/10Problems/index.shtml

Consider two independent random variables, $Z_1 \sim S(\alpha, \beta, 1)$ and $Z_2 \sim S(\alpha, \beta, 1)$. For any non-negative constants $C_1$ and $C_2$, the "$\alpha$-stability" follows from properties of Fourier transforms:

$$Z = C_1 Z_1 + C_2 Z_2 \sim S\left(\alpha, \beta, C_1^\alpha + C_2^\alpha\right).$$

However, if $C_1$ and $C_2$ do not have the same signs, the above "stability" does not hold (unless $\beta = 0$ or $\alpha = 2$, $0+$). To see this, we consider $Z = C_1 Z_1 - C_2 Z_2$, with $C_1 \geq 0$ and $C_2 \geq 0$. Then, because $\mathscr{F}_{-Z_2}(t) = \mathscr{F}_{Z_2}(-t)$,

$$\mathscr{F}_Z = \exp\left(-|C_1 t|^\alpha \left(1 - \sqrt{-1}\beta \mathrm{sgn}(t)\tan\left(\frac{\pi\alpha}{2}\right)\right)\right) \exp\left(-|C_2 t|^\alpha \left(1 + \sqrt{-1}\beta \mathrm{sgn}(t)\tan\left(\frac{\pi\alpha}{2}\right)\right)\right),$$

which does not represent a stable law, unless $\beta = 0$ or $\alpha = 2, 0+$. This is the fundamental reason why *symmetric stable random projections* can be applied to the *general Turnstile* model while our *skewed stable random projections* will be limited to non-negative streams at the time of evaluations. We will soon explain why we recommend $\beta = 1$ (fully skewed).

While there have been numerous studies and applications of random projections, to the best of our knowledge, this is the first proposal for *skewed* stable random projections.

## 1.2 Symmetric Stable Random Projections

Consider a data stream $A_t[i]$, $i \in [1, D]$, following the *Turnstile* model. [11, 12] described the following (idealized) procedure for approximating $F_{(1)} = \sum_{i=1}^{D}(A_t[i])$:

1. Generate $\mathbf{R} \in \mathbb{R}^{D \times k}$ with i.i.d. entries $r_{ij} \sim S(1, 0, 1)$, i.e., standard Cauchy. Set $x_j = 0$, with $j = 1$ to $k$.
2. For each new tuple $a_t = (i, I_t)$, perform $x_j = x_j + I_t \times r_{ij}$, for all $j = 1$ to $k$.
3. Return median($|x_j|, j = 1, ..., k$), as the estimate of $F_{(1)}$.

This procedure extends to $0 < \alpha \leq 2$. By properties of Fourier transforms, the generated $x_j$, $j = 1$ to $k$, represent $k$ i.i.d. samples $x_j \sim S(\alpha, 0, F_{(\alpha)})$. Thus, the problem boils down to estimating the scale parameter $F_{(\alpha)}$ from $k$ i.i.d. samples. The recent paper [16] proposed estimators based on the *geometric mean* and *harmonic mean*.

- The *geometric mean* estimator has variance asymptotically to be $\frac{(\alpha^2+2)\pi^2}{12}\frac{1}{k}F_{(\alpha)}^2$. It exhibits exponential tail bounds and has the sample complexity bound $k = \left(\frac{(\alpha^2+2)\pi^2}{6} + O(\epsilon)\right)\frac{1}{\epsilon^2}\log\left(\frac{2}{\delta}\right)$, so that with probability at least $1 - \delta$, the estimate is within a $1 \pm \epsilon$ factor of the truth.

- The *harmonic mean* estimator is statistically optimal and considerably more accurate than the *geometric mean* estimator, when $\alpha \to 0+$. As $\alpha$ is slightly away from 0, the variance increases substantially and becomes infinite when $\alpha \to 0.5$. This estimator does not have bounds in exponential forms unless $\alpha = 0+$.

## 1.3 Skewed Stable Random Projections

If, at the time $t$ for the evaluation, the data stream is non-negative (which includes the *strict Turnstile* model as a special case), using *symmetric stable random projections* is unnecessary. For example, at $\alpha = 1$, using *symmetric stable random projections* and the *geometric mean* estimator[16], the sample complexity is asymptotically $k = \left(\frac{\pi^2}{2} + O(\epsilon)\right)\frac{1}{\epsilon^2}\log\left(\frac{2}{\delta}\right)$, which is unnecessary, because at $\alpha = 1$, we can use a simple counter to compute $F_{(1)}$ essentially error-free[18, 8, 1, 23]. The problem becomes more interesting when $\alpha$ is slightly larger or smaller than 1. Ideally, we hope to have a mechanism that will be (essentially) error-free when $\alpha \to 1$ in a continuous fashion. The method of *skewed stable random projections* provides such a tool.

Instead of generating the projection matrix $\mathbf{R} \in \mathbb{R}^{D \times k}$ from i.i.d. symmetric stable $r_{ij} \sim S(\alpha, 0, 1)$, we generate $r_{ij} \sim S(\alpha, \beta, 1)$ (and we recommend $\beta = 1$). After the projection operations on the data stream $A_t[i]$, ($i = 1$ to $D$), we obtain $k$ i.i.d. samples $x_j \sim (\alpha, \beta, F_{(\alpha)})$, where $F_{(\alpha)} = \sum_{i=1}^{D}(A_t[i])^\alpha$ is what we are after.

Therefore, we face a new estimation task, which is more sophisticated and less well-studied in statistics than that in *symmetric stable random projections*. Thus, we have to build some of the basic tools from the first statistical principle. We derive the general formula for the moments of skewed stable distributions, based on which we propose the *geometric mean* and *harmonic mean* estimators. In particular, we discover some interesting properties of *fully skewed stable distributions*, which make some estimators have better behaviors (e.g., tail bounds) than previous analogous estimators in [16].

## 1.4 Summary of Estimators

Assume $k$ i.i.d. samples $x_j \sim S\left(\alpha, \beta = 1, F_{(\alpha)}\right)$. We propose five types of estimators and analyze their variances and tail bounds, including the *geometric mean* estimator, the *harmonic mean* estimator, the *maximum likelihood* estimator, as well as the *optimal power* estimator. Figure 1 compares their asymptotic variances along with the asymptotic variance of the *geometric mean* estimator for *symmetric stable random projections*[16].

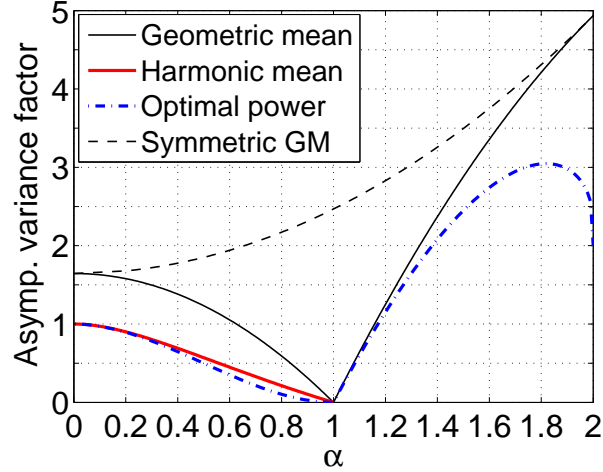

Figure 1: Let $\hat{F}$ be an estimator of $F$ with asymptotic variance $\mathrm{Var}\left(\hat{F}\right) = V\frac{F^2}{k} + O\left(\frac{1}{k^2}\right)$. We plot the $V$ values for the *geometric mean* estimator, the *harmonic mean* estimator (for $\alpha < 1$), the *optimal power* estimator (the lower dashed curve), along with the $V$ values for the *geometric mean* estimator for *symmetric stable random projections* in [16] ("symmetric GM", the upper dashed curve). When $\alpha \to 1$, our method achieves an "infinite improvement" in terms of the asymptotic variances.

### 1.4.1 The geometric mean estimator, $\hat{F}_{(\alpha),gm}$, for $0 < \alpha \leq 2$, ($\alpha \neq 1$)

$$\hat{F}_{(\alpha),gm} = \frac{\prod_{j=1}^{k} |x_j|^{\alpha/k}}{\left(\cos^k\left(\frac{\kappa(\alpha)\pi}{2k}\right)/\cos\left(\frac{\kappa(\alpha)\pi}{2}\right)\right)\left[\frac{2}{\pi}\sin\left(\frac{\pi\alpha}{2k}\right)\Gamma\left(1-\frac{1}{k}\right)\Gamma\left(\frac{\alpha}{k}\right)\right]^k}.$$

$$\mathrm{Var}\left(\hat{F}_{(\alpha),gm}\right) = \frac{F_{(\alpha)}^2}{k}\frac{\pi^2}{12}\left(\alpha^2 + 2 - 3\kappa^2(\alpha)\right) + O\left(\frac{1}{k^2}\right),$$

$$\kappa(\alpha) = \alpha, \text{ if } \alpha < 1, \quad \kappa(\alpha) = 2 - \alpha, \text{ if } \alpha > 1.$$

$\hat{F}_{(\alpha),gm}$ is unbiased and has exponential tail bounds for all $0 < \alpha \leq 2$. We provide the sample complexity bound $k = O\left(G\frac{1}{\epsilon^2}\log\frac{2}{\epsilon}\right)$ explicitly and prove that, as $\alpha = 1 \pm \Delta \to 1$ (i.e., $\Delta \to 0$), for fixed $\epsilon$,

$$G = \frac{\epsilon^2}{\log(1+\epsilon) - 2\sqrt{\Delta\log(1+\epsilon)} + o\left(\sqrt{\Delta}\right)}.$$

### 1.4.2 The harmonic estimator, $\hat{F}_{(\alpha),hm,c}$, for $0 < \alpha < 1$

$$\hat{F}_{(\alpha),hm,c} = \frac{k\frac{\cos\left(\frac{\alpha\pi}{2}\right)}{\Gamma(1+\alpha)}}{\sum_{j=1}^{k}|x_j|^{-\alpha}}\left(1 - \frac{1}{k}\left(\frac{2\Gamma^2(1+\alpha)}{\Gamma(1+2\alpha)} - 1\right)\right),$$

$$\mathrm{E}\left(\hat{F}_{(\alpha),hm,c}\right) = F_{(\alpha)} + O\left(\frac{1}{k^2}\right), \qquad \mathrm{Var}\left(\hat{F}_{(\alpha),hm,c}\right) = \frac{F_{(\alpha)}^2}{k}\left(\frac{2\Gamma^2(1+\alpha)}{\Gamma(1+2\alpha)} - 1\right) + O\left(\frac{1}{k^2}\right).$$

$\hat{F}_{(\alpha),hm,c}$ has exponential tail bounds and we provide the constants explicitly.

### 1.4.3 The maximum likelihood estimator, $\hat{F}_{(0.5),mle,c}$, for $\alpha = 0.5$ only

$$\hat{F}_{(0.5),mle,c} = \left(1 - \frac{3}{4}\frac{1}{k}\right)\sqrt{\frac{k}{\sum_{j=1}^{k}\frac{1}{x_j}}},$$

$$\mathbf{E}\left(\hat{F}_{(0.5),mle,c}\right) = F_{(0.5)} + O\left(\frac{1}{k^2}\right), \qquad \mathrm{Var}\left(\hat{F}_{(0.5),mle,c}\right) = \frac{1}{2}\frac{F_{(0.5)}^2}{k} + \frac{9}{8}\frac{F_{(0.5)}^2}{k^2} + O\left(\frac{1}{k^3}\right).$$

$\hat{F}_{(0.5),mle,c}$ has exponential tail bounds and we provide the constants explicitly.

### 1.4.4 The optimal power estimator, $\hat{F}_{(\alpha),op,c}$, for $0 < \alpha \leq 2$, $(\alpha \neq 1)$

$$\hat{F}_{(\alpha),op,c} = \left(\frac{1}{k}\frac{\sum_{j=1}^{k}|x_j|^{\lambda^*\alpha}}{\frac{\cos(\kappa(\alpha)\frac{\lambda^*\pi}{2})}{\cos^{\lambda^*}\left(\frac{\kappa(\alpha)\pi}{2}\right)}\frac{2}{\pi}\Gamma(1-\lambda^*)\Gamma(\lambda^*\alpha)\sin\left(\frac{\pi}{2}\lambda^*\alpha\right)}\right)^{1/\lambda^*} \times$$
$$\left(1 - \frac{1}{k}\frac{1}{2\lambda^*}\left(\frac{1}{\lambda^*}-1\right)\left(\frac{\cos(\kappa(\alpha)\lambda^*\pi)\frac{2}{\pi}\Gamma(1-2\lambda^*)\Gamma(2\lambda^*\alpha)\sin(\pi\lambda^*\alpha)}{\left[\cos\left(\kappa(\alpha)\frac{\lambda^*\pi}{2}\right)\frac{2}{\pi}\Gamma(1-\lambda^*)\Gamma(\lambda^*\alpha)\sin\left(\frac{\pi}{2}\lambda^*\alpha\right)\right]^2}-1\right)\right),$$

$$\mathbf{E}\left(\hat{F}_{(\alpha),op,c}\right) = F_{(\alpha)} + O\left(\frac{1}{k^2}\right)$$

$$\mathrm{Var}\left(\hat{F}_{(\alpha),op,c}\right) = F_{(\alpha)}^2 \frac{1}{\lambda^{*2}k}\left(\frac{\cos(\kappa(\alpha)\lambda^*\pi)\frac{2}{\pi}\Gamma(1-2\lambda^*)\Gamma(2\lambda^*\alpha)\sin(\pi\lambda^*\alpha)}{\left[\cos\left(\kappa(\alpha)\frac{\lambda^*\pi}{2}\right)\frac{2}{\pi}\Gamma(1-\lambda^*)\Gamma(\lambda^*\alpha)\sin\left(\frac{\pi}{2}\lambda^*\alpha\right)\right]^2}-1\right) + O\left(\frac{1}{k^2}\right).$$

$$\lambda^* = \arg\min g(\lambda;\alpha), \qquad g(\lambda;\alpha) = \frac{1}{\lambda^2}\left(\frac{\cos(\kappa(\alpha)\lambda\pi)\frac{2}{\pi}\Gamma(1-2\lambda)\Gamma(2\lambda\alpha)\sin(\pi\lambda\alpha)}{\left[\cos\left(\kappa(\alpha)\frac{\lambda\pi}{2}\right)\frac{2}{\pi}\Gamma(1-\lambda)\Gamma(\lambda\alpha)\sin\left(\frac{\pi}{2}\lambda\alpha\right)\right]^2}-1\right).$$

When $0 < \alpha < 1$, we prove that $\lambda^* < 0$ and $\hat{F}_{(\alpha),op,c}$ has exponential tail bounds (not explicitly included in the article). $g(\lambda;\alpha)$ is a convex function of $\lambda$, but we provide the rigorous proof only for $0 < \alpha < 1$.

$\hat{F}_{(\alpha),op,c}$ becomes the *harmonic mean* estimator when $\alpha = 0+$, the *arithmetic mean* estimator when $\alpha = 2$, and the *maximum likelihood* estimator when $\alpha = 0.5$.

## 2 The Geometric Mean Estimator

We first prove a fundamental result about the moments of skewed stable distributions.

**Lemma 1** *If $Z \sim S(\alpha, \beta, F_{(\alpha)})$, then for any $\lambda$, where $-1 < \lambda < \alpha$,*

$$\mathbf{E}\left(|Z|^\lambda\right) = F_{(\alpha)}^{\lambda/\alpha}\cos\left(\frac{\lambda}{\alpha}\tan^{-1}\left(\beta\tan\left(\frac{\alpha\pi}{2}\right)\right)\right)\left(1+\beta^2\tan^2\left(\frac{\alpha\pi}{2}\right)\right)^{\frac{\lambda}{2\alpha}}\left(\frac{2}{\pi}\sin\left(\frac{\pi}{2}\lambda\right)\Gamma\left(1-\frac{\lambda}{\alpha}\right)\Gamma(\lambda)\right), \tag{1}$$

*which can be simplified when $\beta = 1$, to be*

$$\mathbf{E}\left(|Z|^\lambda\right) = F_{(\alpha)}^{\lambda/\alpha}\frac{\cos\left(\frac{\kappa(\alpha)}{\alpha}\frac{\lambda\pi}{2}\right)}{\cos^{\lambda/\alpha}\left(\frac{\kappa(\alpha)\pi}{2}\right)}\left(\frac{2}{\pi}\sin\left(\frac{\pi}{2}\lambda\right)\Gamma\left(1-\frac{\lambda}{\alpha}\right)\Gamma(\lambda)\right), \tag{2}$$

$$\kappa(\alpha) = \alpha \quad \text{if} \quad \alpha < 1, \quad \text{and} \quad \kappa(\alpha) = 2-\alpha \quad \text{if} \quad \alpha > 1. \tag{3}$$

For $\alpha < 1$, and $-\infty < \lambda < \alpha$,

$$E\left(|Z|^\lambda\right) = E\left(Z^\lambda\right) = F_{(\alpha)}^{\lambda/\alpha} \frac{\Gamma\left(1 - \frac{\lambda}{\alpha}\right)}{\cos^{\lambda/\alpha}\left(\frac{\alpha\pi}{2}\right)\Gamma(1-\lambda)}. \tag{4}$$

**Proof:** See Appendix A. □

Recall after $k$ projections, we obtain $k$ i.i.d. samples $x_j \sim S(\alpha, \beta, F_{(\alpha)})$ and the task becomes estimating the scale parameter $F_{(\alpha)}$ from these $k$ samples. Setting $\lambda = \frac{\alpha}{k}$ in Lemma 1 yields an unbiased estimator of $F_{(\alpha)}$,

$$\hat{F}_{(\alpha),gm,\beta} = \frac{\prod_{j=1}^k |x_j|^{\alpha/k}}{\cos^k\left(\frac{1}{k}\tan^{-1}\left(\beta\tan\left(\frac{\alpha\pi}{2}\right)\right)\right)\left(1+\beta^2\tan^2\left(\frac{\alpha\pi}{2}\right)\right)^{\frac{1}{2}}\left[\frac{2}{\pi}\sin\left(\frac{\pi\alpha}{2k}\right)\Gamma\left(1-\frac{1}{k}\right)\Gamma\left(\frac{\alpha}{k}\right)\right]^k}. \tag{5}$$

Because of the symmetry about $\beta = 0$, we only consider $0 \leq \beta \leq 1$. In the following Lemma. we show that the variance of $\hat{F}_{(\alpha),gm,\beta}$ decreases with increasing $\beta$.

**Lemma 2** *The variance of $\hat{F}_{(\alpha),gm,\beta}$*

$$\mathrm{Var}\left(\hat{F}_{(\alpha),gm,\beta}\right) = F_{(\alpha)}^2 \left(\frac{\cos^k\left(\frac{2}{k}\tan^{-1}\left(\beta\tan\left(\frac{\alpha\pi}{2}\right)\right)\right)\left[\frac{2}{\pi}\sin\left(\frac{\pi\alpha}{k}\right)\Gamma\left(1-\frac{2}{k}\right)\Gamma\left(\frac{2\alpha}{k}\right)\right]^k}{\cos^{2k}\left(\frac{1}{k}\tan^{-1}\left(\beta\tan\left(\frac{\alpha\pi}{2}\right)\right)\right)\left[\frac{2}{\pi}\sin\left(\frac{\pi\alpha}{2k}\right)\Gamma\left(1-\frac{1}{k}\right)\Gamma\left(\frac{\alpha}{k}\right)\right]^{2k}} - 1\right), \tag{6}$$

*is a decreasing function of $\beta \in [0,1]$.*

**Proof:** It suffices to consider

$$h(\beta) = \frac{\cos\left(\frac{2}{k}\tan^{-1}\left(\beta\tan\left(\frac{\alpha\pi}{2}\right)\right)\right)}{\cos^2\left(\frac{1}{k}\tan^{-1}\left(\beta\tan\left(\frac{\alpha\pi}{2}\right)\right)\right)} = 2 - \sec^2\left(\frac{1}{k}\tan^{-1}\left(\beta\tan\left(\frac{\alpha\pi}{2}\right)\right)\right).$$

*which is a deceasing function of $\beta \in [0,1]$. Thus $\mathrm{Var}\left(\hat{F}_{(\alpha),gm,\beta}\right)$ is also a decreasing function of $\beta \in [0,1]$.* □

Therefore, in order to achieve the smallest variance, we take $\beta = 1$. For brevity, we simply use $\hat{F}_{(\alpha),gm}$ instead of $\hat{F}_{(\alpha),gm,1}$. In fact, for the rest of the paper, we will always consider $\beta = 1$ only.

We rewrite $\hat{F}_{(\alpha),gm}$ (i.e., $\hat{F}_{(\alpha),gm,\beta=1}$) as

$$\hat{F}_{(\alpha),gm} = \frac{\prod_{j=1}^k |x_j|^{\alpha/k}}{\left(\cos^k\left(\frac{\kappa(\alpha)\pi}{2k}\right)/\cos\left(\frac{\kappa(\alpha)\pi}{2}\right)\right)\left[\frac{2}{\pi}\sin\left(\frac{\pi\alpha}{2k}\right)\Gamma\left(1-\frac{1}{k}\right)\Gamma\left(\frac{\alpha}{k}\right)\right]^k}. \tag{7}$$

Recall $\kappa(\alpha) = \alpha$, if $\alpha < 1$, and $\kappa(\alpha) = 2 - \alpha$ if $\alpha > 1$. We need to restrict that $k \geq 2$.

The next Lemma concerns the asymptotic moments of $\hat{F}_{(\alpha),gm}$.

**Lemma 3** *As $k \to \infty$*

$$\left[\cos\left(\frac{\kappa(\alpha)\pi}{2k}\right)\frac{2}{\pi}\Gamma\left(\frac{\alpha}{k}\right)\Gamma\left(1-\frac{1}{k}\right)\sin\left(\frac{\pi}{2}\frac{\alpha}{k}\right)\right]^k \to \exp\left(-\gamma_e(\alpha-1)\right), \tag{8}$$

**monotonically** *with increasing $k$ ($k \geq 2$), where $\gamma_e = 0.57724...$ is Euler's constant.*

*For any fixed $t$, as $k \to \infty$,*

$$E\left(\left(\hat{F}_{(\alpha),gm}\right)^t\right) = F_{(\alpha)}^t \frac{\cos^k\left(\frac{\kappa(\alpha)\pi}{2k}t\right)\left[\frac{2}{\pi}\sin\left(\frac{\pi\alpha}{2k}t\right)\Gamma\left(1-\frac{t}{k}\right)\Gamma\left(\frac{\alpha}{k}t\right)\right]^k}{\cos^{kt}\left(\frac{\kappa(\alpha)\pi}{2k}\right)\left[\frac{2}{\pi}\sin\left(\frac{\pi\alpha}{2k}\right)\Gamma\left(1-\frac{1}{k}\right)\Gamma\left(\frac{\alpha}{k}\right)\right]^{kt}}$$

$$= F_{(\alpha)}^t \exp\left(\frac{1}{k}\frac{\pi^2(t^2-t)}{24}\left(\alpha^2 + 2 - 3\kappa^2(\alpha)\right) + O\left(\frac{1}{k^2}\right)\right). \tag{9}$$

*Consequently,*

$$\mathrm{Var}\left(\hat{F}_{(\alpha),gm}\right) = \frac{F_{(\alpha)}^2}{k}\frac{\pi^2}{12}\left(\alpha^2 + 2 - 3\kappa^2(\alpha)\right) + O\left(\frac{1}{k^2}\right). \tag{10}$$

**Proof:** See Appendix B. □

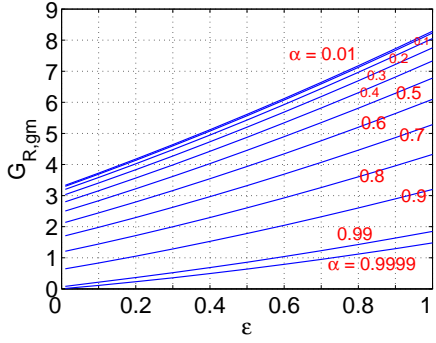
(a) Right tail bound constant, $\alpha < 1$

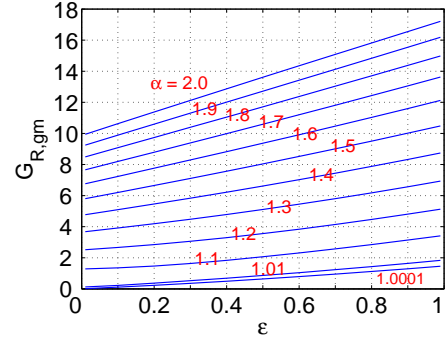
(b) Right tail bound constant, $\alpha > 1$

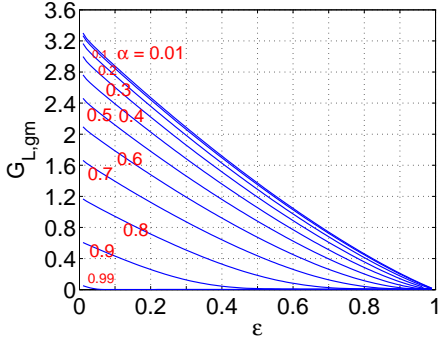
(c) Left tail bound constant, $\alpha < 1$

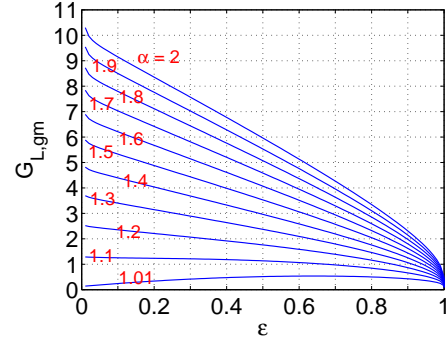
(d) Left tail bound constant, $\alpha > 1$

Figure 2: We plot the tail bound constants of $\hat{F}_{(\alpha),gm}$ in Lemma 4, for a wide range of $\alpha$ and $\epsilon$. For convenience, we plot the left bound constant $G_{L,gm}$ using its asymptote (i.e., assuming $k_0 = \infty$ in (14). This is equivalent to replace the denominator in (7) by its asymptote, which can be viewed as a biased version of the estimator in (7).

Lemma 4 provides the tail bounds and Figure 2 plots the tail bound constants.

**Lemma 4** *The right tail bound:*

$$\mathbf{Pr}\left(\hat{F}_{(\alpha),gm} - F_{(\alpha)} \geq \epsilon F_{(\alpha)}\right) \leq \exp\left(-k\frac{\epsilon^2}{G_{R,gm}}\right), \quad \epsilon > 0. \tag{11}$$

*where*

$$\frac{\epsilon^2}{G_{R,gm}} = C_R \log(1+\epsilon) - C_R \gamma_e (\alpha - 1) - \log\left(\cos\left(\frac{\kappa(\alpha)\pi C_R}{2}\right)\frac{2}{\pi}\Gamma(\alpha C_R)\Gamma(1 - C_R)\sin\left(\frac{\pi \alpha C_R}{2}\right)\right), \tag{12}$$

*and $C_R$ is the solution to*

$$\gamma_e(\alpha - 1) - \log(1+\epsilon) - \frac{\kappa(\alpha)\pi}{2}\tan\left(\frac{\kappa(\alpha)\pi}{2}C_R\right) + \frac{\alpha\pi/2}{\tan\left(\frac{\alpha\pi}{2}C_R\right)} + \psi(\alpha C_R)\alpha - \psi(1 - C_R) = 0.$$

*Here $\psi(z) = \frac{\Gamma'(z)}{\Gamma(z)}$ is the "Psi" function.*
*The left tail bound:*

$$\mathbf{Pr}\left(\hat{F}_{(\alpha),gm} - F_{(\alpha)} \leq -\epsilon F_{(\alpha)}\right) \leq \exp\left(-k\frac{\epsilon^2}{G_{L,gm,k_0}}\right), \quad k > k_0 \;\; 0 < \epsilon < 1. \tag{13}$$

*where*

$$\frac{\epsilon^2}{G_{L,gm,k_0}} = -C_L \log(1-\epsilon) - \log\left(-\cos\left(\frac{\kappa(\alpha)\pi}{2}C_L\right)\frac{2}{\pi}\Gamma(-\alpha C_L)\Gamma(1+C_L)\sin\left(\frac{\pi\alpha C_L}{2}\right)\right)$$
$$- k_0 C_L \log\left(\cos\left(\frac{\kappa(\alpha)\pi}{2k_0}\right)\frac{2}{\pi}\Gamma\left(\frac{\alpha}{k_0}\right)\Gamma\left(1-\frac{1}{k_0}\right)\sin\left(\frac{\pi}{2}\frac{\alpha}{k_0}\right)\right), \quad (14)$$

*and $C_L$ is the solution to*

$$\log(1-\epsilon)C_L - \gamma_e(\alpha-1)C_L + \frac{\kappa(\alpha)\pi}{2}\tan\left(\frac{\kappa(\alpha)\pi}{2}C_L\right) - \frac{\alpha\pi}{2}\tan\left(\frac{\alpha\pi}{2}C_L\right) - \psi(1+\alpha C_L)\alpha + \psi(1+C_L) = 0.$$

***Proof:*** See Appendix C. □

It is interesting and practically important to understand the behavior of the tail bounds when $\alpha = 1 \pm \Delta \to 0$, i.e., $\Delta \to 0$. Figure 3 plots the right tail bound constant $G_{R,gm}$ as a function of $\Delta$ instead of $\alpha$.

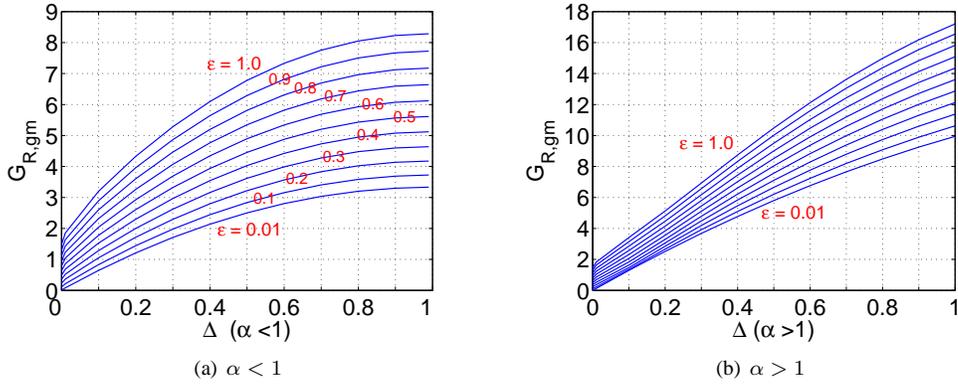

(a) $\alpha < 1$      (b) $\alpha > 1$

Figure 3: We plot the right tail bound constant $G_{R,gm}$ in Lemma 4, as a function of $\Delta$ instead of $\alpha$. Here, we let $0 < \Delta < 1$ always. If $\alpha < 1$, then $\alpha = 1 - \Delta$, and if $\alpha > 1$, then $\alpha = 1 + \Delta$.

Lemma 5 describes the rate of convergence of the right tail bound constant $G_{R,gm}$ as a function of $\Delta$ when $\Delta \to 0$, for fixed $\epsilon$.

**Lemma 5** *Let $\alpha = 1 - \Delta$ if $\alpha < 1$ and $\alpha = 1 + \Delta$ if $\alpha > 1$, i.e. $0 < \Delta < 1$. For fixed $\epsilon$, as $\alpha \to 1$ (i.e., as $\Delta \to 0$), the (right) tail bound constant $G_{R,gm}$ in Lemma 4 converges to $\frac{\epsilon^2}{\log(1+\epsilon)}$ at the rate $O\left(\sqrt{\Delta}\right)$:*

$$G_{R,gm} = \frac{\epsilon^2}{\log(1+\epsilon) - 2\sqrt{\Delta \log(1+\epsilon)} + o\left(\sqrt{\Delta}\right)}. \quad (15)$$

***Proof:*** See Appendix D. □

The fact that $G_{R,gm}$ converges at the rate $O\left(\sqrt{\Delta}\right)$ does not appear completely intuitive. For the sake of verification, Figure 4 plots $G_{R,gm}$ for small values of $\Delta$, along with the approximations suggested in (15).

Once we know the exponential tail bounds, we can establish the sample complexity bound immediately, that $k = O\left(G\frac{1}{\epsilon^2}\log\left(\frac{2}{\delta}\right)\right)$ suffices to approximate $F_{(\alpha)}$ within a $1 \pm \epsilon$ factor with probability at least $1 - \delta$. It suffices to let $G = \max\{G_{R,gm}, G_{L,gm}\}$.

# 3 The Harmonic Mean Estimators for $0 < \alpha < 1$

While the *geometric mean* estimator $\hat{F}_{(\alpha),gm}$ applies to $0 < \alpha \leq 2$ ($\alpha \neq 1$), it is by no means the optimal estimator. For $\alpha < 1$, the *harmonic mean* estimator can considerably improve $\hat{F}_{(\alpha),gm}$. Unlike the *harmonic*

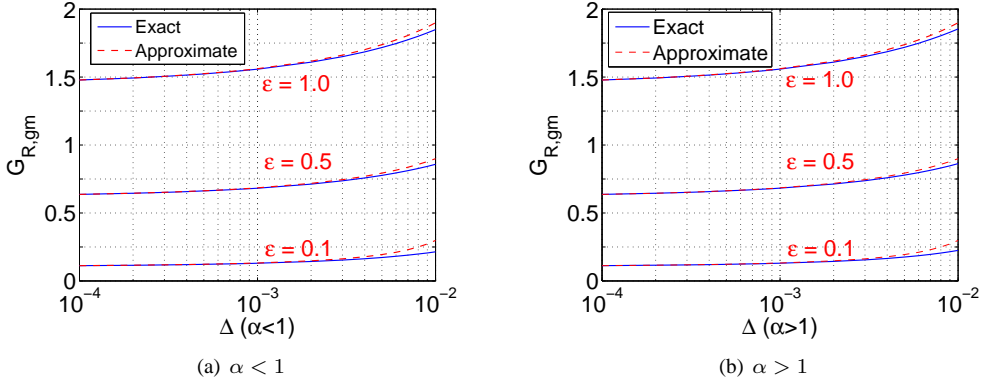

(a) $\alpha < 1$           (b) $\alpha > 1$

Figure 4: We plot $G_{R,gm}$ for small $\Delta$, along with the approximations suggested in (15), i.e., $G_{R,gm} \approx \frac{\epsilon^2}{\log(1+\epsilon) - 2\sqrt{\Delta \log(1+\epsilon)}}$ for small $\Delta$.

*mean* estimator in [16], which is useful only for small $\alpha$ and has no exponential tail bounds except for $\alpha = 0+$, the *harmonic mean* estimator in this study has very nice tail properties for all $0 < \alpha < 1$.

The *harmonic mean* estimator takes advantage of the fact that if $Z \sim S(\alpha < 1, \beta = 1, F_{(\alpha)})$, then $\mathrm{E}\left(|Z|^\lambda\right)$ exists for all $-\infty < \lambda < \alpha$. Note that when $\alpha < 1$ and $\beta = 1$, $Z$ is always non-negative, i.e., $\mathrm{E}\left(|Z|^\lambda\right) = \mathrm{E}\left(Z^\lambda\right)$.

**Lemma 6** *Assume $k$ i.i.d. samples $x_j \sim S(\alpha < 1, \beta = 1, F_{(\alpha)})$, we define the harmonic mean estimator $\hat{F}_{(\alpha),hm}$,*

$$\hat{F}_{(\alpha),hm} = \frac{k \frac{\cos\left(\frac{\alpha\pi}{2}\right)}{\Gamma(1+\alpha)}}{\sum_{j=1}^{k} |x_j|^{-\alpha}}, \tag{16}$$

*and the bias-corrected harmonic mean estimator $\hat{F}_{(\alpha),hm,c}$,*

$$\hat{F}_{(\alpha),hm,c} = \frac{k \frac{\cos\left(\frac{\alpha\pi}{2}\right)}{\Gamma(1+\alpha)}}{\sum_{j=1}^{k} |x_j|^{-\alpha}} \left(1 - \frac{1}{k}\left(\frac{2\Gamma^2(1+\alpha)}{\Gamma(1+2\alpha)} - 1\right)\right). \tag{17}$$

*The bias and variance of $\hat{F}_{(\alpha),hm,c}$ are*

$$E\left(\hat{F}_{(\alpha),hm,c}\right) = F_{(\alpha)} + O\left(\frac{1}{k^2}\right), \tag{18}$$

$$Var\left(\hat{F}_{(\alpha),hm,c}\right) = \frac{F_{(\alpha)}^2}{k}\left(\frac{2\Gamma^2(1+\alpha)}{\Gamma(1+2\alpha)} - 1\right) + O\left(\frac{1}{k^2}\right). \tag{19}$$

*The right tail bound of $\hat{F}_{(\alpha),hm}$ is*

$$\mathbf{Pr}\left(\hat{F}_{(\alpha),hm} - F_{(\alpha)} \geq \epsilon F_{(\alpha)}\right) \leq \exp\left(-k\left(\frac{\epsilon^2}{G_{R,hm}}\right)\right), \qquad \epsilon > 0, \tag{20}$$

$$\frac{\epsilon^2}{G_{R,hm}} = -\log\left(\sum_{m=0}^{\infty} \frac{\Gamma^m(1+\alpha)}{\Gamma(1+m\alpha)}(-t_1^*)^m\right) - \frac{t_1^*}{1+\epsilon}, \tag{21}$$

*where $t_1^*$ is the solution to*

$$\frac{\sum_{m=1}^{\infty}(-1)^m m (t_1^*)^{m-1}\frac{\Gamma^m(1+\alpha)}{\Gamma(1+m\alpha)}}{\sum_{m=0}^{\infty}(-1)^m (t_1^*)^m \frac{\Gamma^m(1+\alpha)}{\Gamma(1+m\alpha)}} + \frac{1}{1+\epsilon} = 0. \tag{22}$$

*The left tail bound of $\hat{F}_{(\alpha),hm}$ is*

$$\mathbf{Pr}\left(\hat{F}_{(\alpha),hm} - F_{(\alpha)} \leq -\epsilon F_{(\alpha)}\right) \leq \exp\left(-k\left(\frac{\epsilon^2}{G_{L,hm}}\right)\right), \quad 0 < \epsilon < 1, \quad (23)$$

$$\frac{\epsilon^2}{G_{L,hm}} = -\log\left(\sum_{m=0}^{\infty} \frac{\Gamma^m(1+\alpha)}{\Gamma(1+m\alpha)}(t_2^*)^m\right) + \frac{t_2^*}{1-\epsilon} \quad (24)$$

*where $t_2^*$ is the solution to*

$$-\frac{\sum_{m=1}^{\infty} m(t_2^*)^{m-1} \frac{\Gamma^m(1+\alpha)}{\Gamma(1+m\alpha)}}{\sum_{m=0}^{\infty} (t_2^*)^m \frac{\Gamma^m(1+\alpha)}{\Gamma(1+m\alpha)}} + \frac{1}{1-\epsilon} = 0 \quad (25)$$

***Proof:*** See Appendix E. □.

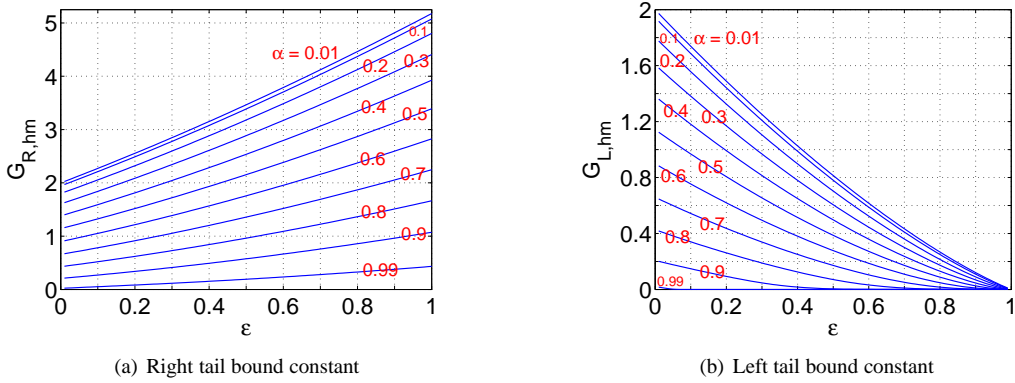

(a) Right tail bound constant  (b) Left tail bound constant

Figure 5: We plot the tail bound constants for the *harmonic mean* estimator in Lemma 6.

## 4 The Maximum Likelihood Estimators for $\alpha = 0.5$

Estimators based on the maximum likelihood are statistically optimal (though usually biased). It is known that the optimal estimator for $F_{(2)}$ is the *arithmetic mean*, which is the maximum likelihood estimator (MLE). [16] has shown that the *harmonic mean* estimator is the MLE for $\alpha = 0+$. This section analyzes the MLE for $\alpha = 0.5$, which corresponds to the *Lévy* distribution. Suppose $X \sim S(\alpha = 0.5, \beta = 1, F_{(0.5)})$. Then

$$f_Z(z) = \frac{F_{(0.5)}}{\sqrt{2\pi}} \frac{\exp\left(-\frac{F_{(0.5)}^2}{2z}\right)}{z^{3/2}}, \qquad F_Z(z) = \frac{2}{\sqrt{\pi}} \int_{\sqrt{\frac{1}{2z}}}^{\infty} e^{-t^2} dt = \operatorname{erfc}\left(\sqrt{\frac{1}{2z}}\right). \quad (26)$$

The next Lemma derives the maximum likelihood estimators and their moments.

**Lemma 7** *Assume $k$ i.i.d. samples $x_j \sim S(0.5, 1, F_{(0.5)})$, the maximum likelihood estimator of $F_{(0.5)}$, is*

$$\hat{F}_{(0.5),mle} = \sqrt{\frac{k}{\sum_{j=1}^{k} \frac{1}{x_j}}}. \quad (27)$$

*To reduce the bias and variance, we recommend the bias-corrected version:*

$$\hat{F}_{(0.5),mle,c} = \left(1 - \frac{3}{4}\frac{1}{k}\right)\hat{F}_{(0.5),mle} = \left(1 - \frac{3}{4}\frac{1}{k}\right)\sqrt{\frac{k}{\sum_{j=1}^{k} \frac{1}{x_j}}}. \quad (28)$$

*The first four moments of $\hat{h}_{mle,c}$ are*

$$E\left(\hat{F}_{(0.5),mle,c}\right) = F_{(0.5)} + O\left(\frac{1}{k^2}\right), \tag{29}$$

$$\text{Var}\left(\hat{F}_{(0.5),mle,c}\right) = \frac{1}{2}\frac{F_{(0.5)}^2}{k} + \frac{9}{8}\frac{F_{(0.5)}^2}{k^2} + O\left(\frac{1}{k^3}\right), \tag{30}$$

$$E\left(\hat{F}_{(0.5),mle,c} - E\left(\hat{F}_{(0.5),mle,c}\right)\right)^3 = \frac{5}{4}\frac{F_{(0.5)}^3}{k^2} + O\left(\frac{1}{k^3}\right), \tag{31}$$

$$E\left(\hat{F}_{(0.5),mle,c} - E\left(\hat{F}_{(0.5),mle,c}\right)\right)^4 = \frac{3}{4}\frac{F_{(0.5)}^4}{k^2} + \frac{75}{8}\frac{F_{(0.5)}^4}{k^3} + O\left(\frac{1}{k^4}\right). \tag{32}$$

*Proof: See Appendix F.* □.

Compared with the *geometric mean estimator* at $\alpha = 0.5$, whose variance is $1.2337\frac{F_{(0.5)}^2}{k} + O\left(\frac{1}{k^2}\right)$, we can see that $\hat{F}_{(0.5),mle,c}$ significantly reduces the variance. Compared with the *harmonic mean* estimator at $\alpha = 0.5$, whose variance is $\frac{0.5708}{k}F_{(0.5)}^2 + O\left(\frac{1}{k^2}\right)$, the variance of $\hat{F}_{(0.5),mle,c}$ is still smaller.

The next task is to derive tail bounds. Although we recommend the bias-corrected version $\hat{F}_{(0.5),mle,c}$, for convenience, we actually present the tail bounds only for $\hat{F}_{(0.5),mle}$.

**Lemma 8**

$$\mathbf{Pr}\left(\hat{F}_{(0.5),mle} - F_{(0.5)} \geq \epsilon F_{(0.5)}\right) \leq \exp\left(-k\left(\log(1+\epsilon) - \frac{1}{2} + \frac{1}{2}\frac{1}{(1+\epsilon)^2}\right)\right), \quad \epsilon > 0, \tag{33}$$

$$\mathbf{Pr}\left(\hat{F}_{(0.5),mle} - F_{(0.5)} \leq -\epsilon F_{(0.5)}\right) \leq \exp\left(-k\left(\log(1-\epsilon) - \frac{1}{2} + \frac{1}{2}\frac{1}{(1-\epsilon)^2}\right)\right), \quad 0 < \epsilon < 1. \tag{34}$$

*For small $\epsilon$, the tail bounds can be written as*

$$\mathbf{Pr}\left(\hat{F}_{(0.5),mle} - F_{(0.5)} \geq \epsilon F_{(0.5)}\right) \leq \exp\left(-k\left(\epsilon^2 - \frac{5}{3}\epsilon^3 + ...\right)\right), \tag{35}$$

$$\mathbf{Pr}\left(\hat{F}_{(0.5),mle} - F_{(0.5)} \leq -\epsilon F_{(0.5)}\right) \leq \exp\left(-k\left(\epsilon^2 + \frac{5}{3}\epsilon^3 + ...\right)\right). \tag{36}$$

*Proof: See Appendix G.* □.

## 5 The Optimal Power Estimator

One may have noticed that, the MLE at $\alpha = 0.5$, the *harmonic mean* estimator at $\alpha = 0+$, and the *arithmetic mean* estimator for $\alpha = 2$, share the same *fractional power* form. Thus, this section is devoted to the *optimal power* estimator.

**Lemma 9** *The* optimal power *estimator:*

$$\hat{F}_{(\alpha),op,c} = \left(\frac{1}{k}\frac{\sum_{j=1}^{k}|x_j|^{\lambda^*\alpha}}{\frac{\cos\left(\kappa(\alpha)\frac{\lambda^*\pi}{2}\right)}{\cos^{\lambda^*}\left(\frac{\kappa(\alpha)\pi}{2}\right)}\frac{2}{\pi}\Gamma(1-\lambda^*)\Gamma(\lambda^*\alpha)\sin\left(\frac{\pi}{2}\lambda^*\alpha\right)}\right)^{1/\lambda^*} \times$$

$$\left(1 - \frac{1}{k}\frac{1}{2\lambda^*}\left(\frac{1}{\lambda^*}-1\right)\left(\frac{\cos\left(\kappa(\alpha)\lambda^*\pi\right)\frac{2}{\pi}\Gamma(1-2\lambda^*)\Gamma(2\lambda^*\alpha)\sin\left(\pi\lambda^*\alpha\right)}{\left[\cos\left(\kappa(\alpha)\frac{\lambda^*\pi}{2}\right)\frac{2}{\pi}\Gamma(1-\lambda^*)\Gamma(\lambda^*\alpha)\sin\left(\frac{\pi}{2}\lambda^*\alpha\right)\right]^2} - 1\right)\right), \tag{37}$$

*has bias and variance*

$$E\left(\hat{F}_{(\alpha),op,c}\right) = F_{(\alpha)} + O\left(\frac{1}{k^2}\right) \tag{38}$$

$$Var\left(\hat{F}_{(\alpha),op,c}\right) = F_{(\alpha)}^2 \frac{1}{\lambda^{*2}k}\left(\frac{\cos\left(\kappa(\alpha)\lambda^*\pi\right)\frac{2}{\pi}\Gamma(1-2\lambda^*)\Gamma(2\lambda^*\alpha)\sin\left(\pi\lambda^*\alpha\right)}{\left[\cos\left(\kappa(\alpha)\frac{\lambda^*\pi}{2}\right)\frac{2}{\pi}\Gamma(1-\lambda^*)\Gamma(\lambda^*\alpha)\sin\left(\frac{\pi}{2}\lambda^*\alpha\right)\right]^2} - 1\right) + O\left(\frac{1}{k^2}\right). \tag{39}$$

*where*

$$\lambda^* = argmin\ g\left(\lambda;\alpha\right),\quad g\left(\lambda;\alpha\right) = \frac{1}{\lambda^2}\left(\frac{\cos\left(\kappa(\alpha)\lambda\pi\right)\frac{2}{\pi}\Gamma(1-2\lambda)\Gamma(2\lambda\alpha)\sin\left(\pi\lambda\alpha\right)}{\left[\cos\left(\kappa(\alpha)\frac{\lambda\pi}{2}\right)\frac{2}{\pi}\Gamma(1-\lambda)\Gamma(\lambda\alpha)\sin\left(\frac{\pi}{2}\lambda\alpha\right)\right]^2} - 1\right). \tag{40}$$

***Proof:*** *See Appendix H.*□

Figure 6(a) plots $g(\lambda;\alpha)$ in Lemma 9 as functions of $\lambda$ for a good range of $\alpha$ values, illustrating that $g(\lambda;\alpha)$ is a convex function of $\lambda$ and hence the minimums $\lambda^*$ can be easily obtained. Figure 6(b) plots the optimal values $\lambda^*$ a function of $\alpha$.

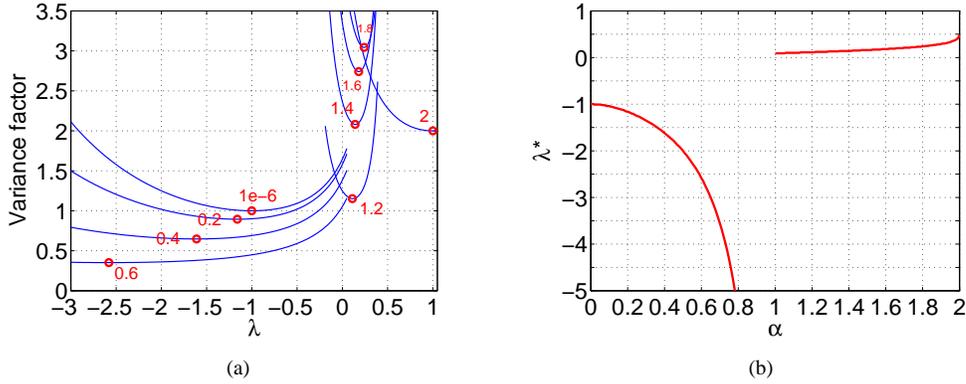

Figure 6: (a)We plot $g(\lambda;\alpha)$ in Lemma 9 as functions of $\lambda$ for a good range of $\alpha$ values, illustrating that $g(\lambda;\alpha)$ is a convex function of $\lambda$ and hence the minimums $\lambda^*$ can be easily obtained (i.e., the lowest points on the curves). Note that there is a singularity at $\alpha = 2-$. (b) We plot the optimal values $\lambda^*$ a function of $\alpha$, only for $0 < \alpha < 2$.

This type of estimator was recently proposed in [17], for *symmetric stable random projections*, by aggressively minimizing the asymptotic variance from the solution to a convex program. The problem with the *fractional power* estimator in [17] is that it only has finite moments to a rather limited order (which seriously affect tail behaviors).

The story is somewhat different for the *fractional power* estimator in this section, although the analysis becomes more complicated than in [17]. For $\alpha < 1$, Lemma 10 proves that the optimal power $\lambda^* < 0$, implying that all moments exist and exponential tail bounds hold. Lemma 10 also proves that $g(\lambda;\alpha)$ is a convex function of $\lambda$.

**Lemma 10** *If $\alpha < 1$, then $g(\lambda;\alpha)$ is a convex function of $\lambda$ and the optimal solution $\lambda^* < 0$.*
***Proof:*** *See Appendix I.*□

The fact that $\lambda^* < 0$ when $\alpha < 1$ is very useful, because it implies that the estimator has all the moments when $\alpha < 1$ and consequently exponential tail bounds exist.

When $\alpha = 0.5$, we can verify that $\lambda = -2$ satisfies $\frac{\partial g(\lambda;\alpha)}{\partial \lambda} = 0$. Because $g(\lambda;\alpha)$ is a convex function, we know $\lambda^* = -2$ when $\alpha = 0.5$, and $\hat{F}_{(0.5),op,c}$ is exactly the maximum likelihood estimator at $\alpha = 0.5$, i.e.,

$$\hat{F}_{(0.5),op,c} = \left(1 - \frac{3}{4}\frac{1}{k}\right)\sqrt{\frac{k}{\sum_{j=1}^{k}\frac{1}{x_j}}}.$$

Therefore, the *optimal power* estimator becomes statistically optimal at least at $\alpha = 0+$, $\alpha = 2$, and $\alpha = 0.5$.

# 6  Conclusion

Approximating the $\alpha$th frequency moments in massive data streams is a frequently studied problem. In some applications, we might treat $\alpha$ as a *tuning* parameter. In other applications, $\alpha$ may bear some physical meaning, for example, $\alpha = 1 \pm \Delta$ with $\Delta$ being the "decay rate" or "interest rate," where $\Delta$ is often small.

We consider the popular *Turnstile* data stream model, which allows both insertions and deletions. We propose a new method called *skewed stable random projections* for approximating the $\alpha$th frequency moments (where $0 < \alpha \leq 2$) on data streams that are: (a) insertion only (i.e., *cash register* model), or (b) always non-negative (i.e., *strict Turnstile* model), or (c) eventually non-negative at check points. Because of the natural constraints in real-world, we believe our model suffices for describing most data streams encountered in practice.

Our proposed method works particulary well when $\alpha$ is about 1, which correspond to many practical settings. For example, we can view *skewed stable random projections* as a "generalized counter" for approximating the total values in the future taking into account the effect of decaying or interest accruement.

In this paper, detailed statistical analysis is conducted on a variety of estimators derived from the first principle, including estimators based on the *geometric mean*, the *harmonic mean*, the *maximum likelihood*, and the *fractional power*. The *geometric mean* estimator is particularly useful for theoretical analysis of the sample complexity bound as well as the local behavior of the sample complexity when $\alpha \to 1$. For example, we show that using the *geometric mean* estimator, the sample complexity bound constant converges to $\epsilon^2/\log(1+\epsilon)$ when $\alpha = 1 \pm \Delta \to 1$, at the rate $O\left(\sqrt{\Delta}\right)$.

To conclude the paper, we should mention that in some applications, *skewed stable random projections* may be combined with *symmetric stable random projections*, due to the linearity in the definition of the $\alpha$th frequency moment. For example, we can use *skewed stable random projections* for those elements which we are certain that they will eventually turn non-negative at least at the time of evaluations; and we can use *symmetric stable random projections* for those elements which we are less certain about the signs.

# 7  Acknowledgement

The author wishes to thank some helpful discussions and suggestions from Gennady Samorodnitsky, Jon Kleinberg, Martin Wells, and Anand Vidyashankar. The author appreciates that Jelani Nelson (and the research group at MIT) mentioned some immediate applications of *skewed stable random projections* using $\alpha$ very close 1, after the author presented some of the unpublished results in this paper at SODA'08.

# A  Proof of Lemma 1

Assume $Z \sim S(\alpha, \beta, F_{(\alpha)})$. To prove $\mathbf{E}\left(|Z|^\lambda\right)$ for $-1 < \lambda < \alpha$, [26, Theorem 2.6.3] provided a partial answer:

$$\int_0^\infty z^\lambda f_Z(z; \alpha, \beta_B, F_{(\alpha)}) dz = F_{(\alpha)}^{\lambda/\alpha} \frac{\sin(\pi\rho\lambda)}{\sin(\pi\lambda)} \frac{\Gamma\left(1 - \frac{\lambda}{\alpha}\right)}{\Gamma(1-\lambda)} \cos^{-\lambda/\alpha}(\pi\beta_B \kappa(\alpha)/2)$$

where we denote

$$\kappa(\alpha) = \alpha \text{ if } \alpha < 1, \text{ and } \kappa(\alpha) = 2 - \alpha \text{ if } \alpha > 1,$$

and according to the notation and parametrization in the book[26, I.19, I.28] :

$$\beta_B = \frac{2}{\pi\kappa(\alpha)} \tan^{-1}\left(\beta \tan\left(\frac{\pi\alpha}{2}\right)\right), \quad \rho = \frac{1 - \beta_B \kappa(a)/\alpha}{2}.$$

Note that

$$\cos^{-\lambda/\alpha}(\pi\beta_B \kappa(\alpha)/2) = \left(1 + \tan^2(\pi\beta_B\kappa(\alpha)/2)\right)^{\frac{\lambda}{2\alpha}}$$

$$= \left(1 + \tan^2\left(\tan^{-1}\left(\beta\tan\left(\frac{\pi\alpha}{2}\right)\right)\right)\right)^{\frac{\lambda}{2\alpha}} = \left(1 + \beta^2 \tan^2\left(\frac{\pi\alpha}{2}\right)\right)^{\frac{\lambda}{2\alpha}}.$$

Therefore, for $-1 < \lambda < \alpha$, [26, Theorem 2.6.3] is equivalent to

$$\int_0^\infty z^\lambda f_Z(z; \alpha, \beta_B, F_{(\alpha)}) dz = F_{(\alpha)}^{\lambda/\alpha} \frac{\sin(\pi \rho \lambda)}{\sin(\pi \lambda)} \frac{\Gamma\left(1 - \frac{\lambda}{\alpha}\right)}{\Gamma(1 - \lambda)} \left(1 + \beta^2 \tan^2\left(\frac{\pi\alpha}{2}\right)\right)^{\frac{\lambda}{2\alpha}}.$$

To compute $\mathrm{E}\left(|Z|^\lambda\right)$, we take advantage of a useful property of the stable density function[26, page 65]:

$$f_Z(-z; \alpha, \beta_B, F_{(\alpha)}) = f_Z(z; \alpha, -\beta_B, F_{(\alpha)}).$$

$$\begin{aligned}
\mathrm{E}\left(|Z|^\lambda\right) &= \int_{-\infty}^0 (-z)^\lambda f_Z(z; \alpha, \beta_B, F_{(\alpha)}) dz + \int_0^\infty z^\lambda f_Z(z; \alpha, \beta_B, F_{(\alpha)}) dz \\
&= \int_0^\infty z^\lambda f_Z(z; \alpha, -\beta_B, F_{(\alpha)}) dz + \int_0^\infty z^\lambda f_Z(z; \alpha, \beta_B, F_{(\alpha)}) dz \\
&= \frac{F_{(\alpha)}^{\lambda/\alpha}}{\sin(\pi\lambda)} \frac{\Gamma\left(1 - \frac{\lambda}{\alpha}\right)}{\Gamma(1 - \lambda)} \left(1 + \beta^2 \tan^2\left(\frac{\pi\alpha}{2}\right)\right)^{\frac{\lambda}{2\alpha}} \left(\sin\left(\pi\lambda \frac{1 - \beta_B \kappa(\alpha)/\alpha}{2}\right) + \sin\left(\pi\lambda \frac{1 + \beta_B \kappa(\alpha)/\alpha}{2}\right)\right) \\
&= \frac{F_{(\alpha)}^{\lambda/\alpha}}{\sin(\pi\lambda)} \frac{\Gamma\left(1 - \frac{\lambda}{\alpha}\right)}{\Gamma(1 - \lambda)} \left(1 + \beta^2 \tan^2\left(\frac{\pi\alpha}{2}\right)\right)^{\frac{\lambda}{2\alpha}} \left(2 \sin\left(\frac{\pi\lambda}{2}\right) \cos\left(\frac{\pi\lambda}{2} \beta_B \kappa(\alpha)/\alpha\right)\right) \\
&= \frac{F_{(\alpha)}^{\lambda/\alpha}}{\cos(\pi\lambda/2)} \frac{\Gamma\left(1 - \frac{\lambda}{\alpha}\right)}{\Gamma(1 - \lambda)} \left(1 + \beta^2 \tan^2\left(\frac{\pi\alpha}{2}\right)\right)^{\frac{\lambda}{2\alpha}} \cos\left(\frac{\lambda}{\alpha} \tan^{-1}\left(\beta \tan\left(\frac{\pi\alpha}{2}\right)\right)\right) \\
&= F_{(\alpha)}^{\lambda/\alpha} \left(1 + \beta^2 \tan^2\left(\frac{\pi\alpha}{2}\right)\right)^{\frac{\lambda}{2\alpha}} \cos\left(\frac{\lambda}{\alpha} \tan^{-1}\left(\beta \tan\left(\frac{\pi\alpha}{2}\right)\right)\right) \left(\frac{2}{\pi} \sin\left(\frac{\pi}{2}\lambda\right) \Gamma\left(1 - \frac{\lambda}{\alpha}\right) \Gamma(\lambda)\right),
\end{aligned}$$

which can be simplified when $\beta = 1$, to be

$$\mathrm{E}\left(|Z|^\lambda\right) = F_{(\alpha)}^{\lambda/\alpha} \frac{\cos\left(\frac{\kappa(\alpha)}{\alpha} \frac{\lambda\pi}{2}\right)}{\cos^{\lambda/\alpha}\left(\frac{\kappa(\alpha)\pi}{2}\right)} \left(\frac{2}{\pi} \sin\left(\frac{\pi}{2}\lambda\right) \Gamma\left(1 - \frac{\lambda}{\alpha}\right) \Gamma(\lambda)\right).$$

The final task is to show that when $\alpha < 1$ and $\beta = 1$, $\mathrm{E}\left(|Z|^\lambda\right)$ exists for all $-\infty < \lambda < \alpha$, not just $-1 < \lambda < \alpha$. This is an extremely useful property.

Note that when $\alpha < 1$ and $\beta = 1$, $Z$ is always non-negative. As shown in the proof of [26, Theorem 2.6.3],

$$\begin{aligned}
\mathrm{E}\left(|Z|^\lambda\right) &= F_{(\alpha)}^{\lambda/\alpha} \cos^{-\lambda/\alpha}\left(\frac{\pi\alpha}{2}\right) \frac{1}{\pi} \mathrm{Im} \int_0^\infty z^\lambda \int_0^\infty \exp\left(-zu \exp(\sqrt{-1}\pi/2) - u^\alpha \exp(-\sqrt{-1}\pi\alpha/2) + \frac{\sqrt{-1}\pi}{2}\right) du dz \\
&= F_{(\alpha)}^{\lambda/\alpha} \cos^{-\lambda/\alpha}\left(\frac{\pi\alpha}{2}\right) \frac{1}{\pi} \mathrm{Im} \int_0^\infty \int_0^\infty z^\lambda \exp\left(-zu\sqrt{-1} - u^\alpha \exp(-\sqrt{-1}\pi\alpha/2)\right) \sqrt{-1} du dz.
\end{aligned}$$

The only thing we need to check is that in the proof of [26, Theorem 2.6.3], the condition for Fubini's theorem (to exchange order of integration) still holds when $-\infty < \alpha < 1$, $\beta = 1$, and $\lambda < -1$. We can show

$$\begin{aligned}
\int_0^\infty \int_0^\infty & \left|z^\lambda \exp\left(-zu\sqrt{-1} - u^\alpha \exp(-\sqrt{-1}\pi\alpha/2)\right) \sqrt{-1}\right| du dz \\
&= \int_0^\infty \int_0^\infty z^\lambda \left|\exp\left(-u^\alpha \cos(\pi\alpha/2) + \sqrt{-1} u^\alpha \sin(\pi\alpha/2)\right)\right| du dz \\
&= \int_0^\infty \int_0^\infty z^\lambda \exp\left(-u^\alpha \cos(\pi\alpha/2)\right) du dz < \infty,
\end{aligned}$$

provided $\lambda < -1$ ($\lambda \neq -1, -2, -3, \ldots$) and $\cos(\pi\alpha/2) > 0$, i.e., $\alpha < 1$. Note that $|\exp(\sqrt{-1}x)| = 1$ always and Euler's formula: $\exp(\sqrt{-1}x) = \cos(x) + \sqrt{-1}\sin(x)$ is frequently used to simplify the algebra.

Once we have shown that Fubini's condition is satisfied, we can exchange the order of integration and the rest just follows from the proof of [26, Theorem 2.6.3]. Note that because of continuity, the "singularity points" $\lambda = -1, -2, -3, \ldots$ do not matter.

We should mention that in an unpublished technical report[14], cited as [21, Property 1.2.17]), $\mathrm{E}\left(|Z|^\lambda\right)$ was proved in an integral form, but only for $0 < \lambda < \alpha$.

# B  Proof of Lemma 3

We first show that, for any fixed $t$, as $k \to \infty$,

$$\mathrm{E}\left(\left(\hat{F}_{(\alpha),gm}\right)^t\right) = F_{(\alpha)}^t \frac{\cos^k\left(\frac{\kappa(\alpha)\pi}{2k}t\right)\left[\frac{2}{\pi}\sin\left(\frac{\pi\alpha}{2k}t\right)\Gamma\left(1-\frac{t}{k}\right)\Gamma\left(\frac{\alpha}{k}t\right)\right]^k}{\cos^{kt}\left(\frac{\kappa(\alpha)\pi}{2k}\right)\left[\frac{2}{\pi}\sin\left(\frac{\pi\alpha}{2k}\right)\Gamma\left(1-\frac{1}{k}\right)\Gamma\left(\frac{\alpha}{k}\right)\right]^{kt}}$$

$$= F_{(\alpha)}^t \exp\left(\frac{1}{k}\frac{\pi^2(t^2-t)}{24}\left(\alpha^2 + 2 - 3\kappa^2(\alpha)\right) + O\left(\frac{1}{k^2}\right)\right).$$

In [16], it was proved that, as $k \to \infty$,

$$\frac{\left[\frac{2}{\pi}\sin\left(\frac{\pi\alpha}{2k}t\right)\Gamma\left(1-\frac{t}{k}\right)\Gamma\left(\frac{\alpha}{k}t\right)\right]^k}{\left[\frac{2}{\pi}\sin\left(\frac{\pi\alpha}{2k}\right)\Gamma\left(1-\frac{1}{k}\right)\Gamma\left(\frac{\alpha}{k}\right)\right]^{kt}} = 1 + \frac{1}{k}\frac{\pi^2(t^2-t)}{24}(\alpha^2+2) + O\left(\frac{1}{k^2}\right)$$

$$= \exp\left(\frac{1}{k}\frac{\pi^2(t^2-t)}{24}(\alpha^2+2) + O\left(\frac{1}{k^2}\right)\right).$$

Using the infinite product representation of the $\cos$ function[9, 1.43.3]

$$\cos(z) = \prod_{s=0}^{\infty}\left(1 - \frac{4z^2}{(2s+1)^2\pi^2}\right),$$

we can rewrite

$$\frac{\cos^k\left(\frac{\kappa(\alpha)\pi}{2k}t\right)}{\cos^{kt}\left(\frac{\kappa(\alpha)\pi}{2k}\right)} = \prod_{s=0}^{\infty}\left(1 - \frac{\kappa^2(\alpha)t^2}{(2s+1)^2 k^2}\right)^k \left(1 - \frac{\kappa^2(\alpha)}{(2s+1)^2 k^2}\right)^{-kt}$$

$$= \prod_{s=0}^{\infty}\left(\left(1 - \frac{\kappa^2(\alpha)t^2}{(2s+1)^2 k^2}\right)\left(1 + t\frac{\kappa^2(\alpha)}{(2s+1)^2 k^2} + O\left(\frac{1}{k^3}\right)\right)\right)^k$$

$$= \prod_{s=0}^{\infty}\left(1 - \frac{\kappa^2(\alpha)(t^2-t)}{(2s+1)^2 k^2} + O\left(\frac{1}{k^3}\right)\right)^k = \prod_{s=0}^{\infty}\left(1 - \frac{\kappa^2(\alpha)(t^2-t)}{(2s+1)^2 k} + O\left(\frac{1}{k^2}\right)\right)$$

$$= \exp\left(\sum_{s=0}^{\infty}\log\left(1 - \frac{\kappa^2(\alpha)(t^2-t)}{(2s+1)^2 k} + O\left(\frac{1}{k^2}\right)\right)\right)$$

$$= \exp\left(-\frac{\kappa^2(\alpha)}{k}(t^2-t)\sum_{s=0}^{\infty}\frac{1}{(2s+1)^2} + O\left(\frac{1}{k^2}\right)\right)$$

$$= \exp\left(-\frac{\kappa^2(\alpha)}{k}(t^2-t)\frac{\pi^2}{8} + O\left(\frac{1}{k^2}\right)\right),$$

which, combined with the result in [16], yields the desired expression.

The next task is to show

$$\left[\cos\left(\frac{\kappa(\alpha)\pi}{2k}\right)\frac{2}{\pi}\Gamma\left(\frac{\alpha}{k}\right)\Gamma\left(1-\frac{1}{k}\right)\sin\left(\frac{\pi}{2}\frac{\alpha}{k}\right)\right]^k \to \exp\left(-\gamma_e(\alpha-1)\right),$$

monotonically as $k \to \infty$, where $\gamma_e = 0.577215665...$, is Euler's constant.

In [16], it was proved that, as $k \to \infty$,

$$\left[\frac{2}{\pi}\Gamma\left(\frac{\alpha}{k}\right)\Gamma\left(1-\frac{1}{k}\right)\sin\left(\frac{\pi}{2}\frac{\alpha}{k}\right)\right]^k \to \exp\left(-\gamma_e(\alpha-1)\right),$$

monotonically. In this study, we need to consider instead

$$\left[\cos\left(\frac{\kappa(\alpha)\pi}{2k}\right)\frac{2}{\pi}\Gamma\left(\frac{\alpha}{k}\right)\Gamma\left(1-\frac{1}{k}\right)\sin\left(\frac{\pi}{2}\frac{\alpha}{k}\right)\right]^k = \left[2\cos\left(\frac{\kappa(\alpha)\pi}{2k}\right)\frac{\Gamma\left(\frac{\alpha}{k}\right)\sin\left(\frac{\pi\alpha}{2k}\right)}{\Gamma\left(\frac{1}{k}\right)\sin\left(\frac{\pi}{k}\right)}\right]^k \quad (41)$$

Note that the additional term $\left[\cos\left(\frac{\kappa(\alpha)\pi}{2k}\right)\right]^k = 1 - O\left(\frac{1}{k}\right)$. Therefore,

$$\left[\cos\left(\frac{\kappa(\alpha)\pi}{2k}\right)\frac{2}{\pi}\Gamma\left(\frac{\alpha}{k}\right)\Gamma\left(1-\frac{1}{k}\right)\sin\left(\frac{\pi}{2}\frac{\alpha}{k}\right)\right]^k \to \exp\left(-\gamma_e\left(\alpha-1\right)\right).$$

To show the monotonicity, however, we have to use some different techniques from [16]. The reason is because the additional term $\left[\cos\left(\frac{\kappa(\alpha)\pi}{2k}\right)\right]^k$ increases (instead of decreasing) monotonically with increasing $k$.

First, we consider $\alpha > 1$, i.e., $\kappa(\alpha) = 2 - \alpha < 1$. For simplicity, we take logarithm of (41) and replace $1/k$ by $t$, where $0 \leq t \leq 1/2$ (recall $k \geq 2$). It suffices to show that $g(t)$ increases with increasing $t \in [0, 1/2]$, where

$$g(t) = \frac{1}{t}W(t),$$

$$W(t) = \log\left(\cos\left(\frac{\kappa(\alpha)\pi}{2}t\right)\right) + \log\left(\Gamma\left(\alpha t\right)\right) + \log\left(\sin\left(\frac{\pi\alpha}{2}t\right)\right) - \log\left(\Gamma\left(t\right)\right) - \log\left(\sin\left(\pi t\right)\right) + \log(2).$$

Because $g'(t) = \frac{1}{t}W'(t) - \frac{1}{t^2}W(t)$, to show $g'(t) \geq 0$ in $t \in [0, 1/2]$, it suffices to show

$$tW'(t) - W(t) \geq 0.$$

One can check that $tW'(t) \to 0$ and $W(t) \to 0$, as $t \to 0+$, where

$$W'(t) = -\tan\left(\frac{\kappa(\alpha)\pi}{2}t\right)\left(\frac{\kappa\pi}{2}\right) + \psi\left(\alpha t\right)\alpha + \frac{1}{\tan\left(\frac{\pi\alpha}{2}t\right)}\left(\frac{\alpha\pi}{2}\right) - \psi(t) - \frac{1}{\tan\left(\pi t\right)}\pi.$$

Here $\psi(x) = \frac{\partial \log(\Gamma(x))}{\partial x}$ is the "Psi" function.

Therefore, to show $tW'(t) - W(t) \geq 0$, it suffices to show that $tW'(t) - W(t)$ is an increasing function of $t \in [0, 1/2]$, i.e.,

$$(tW'(t) - W(t))' = W''(t) \geq 0, \quad \text{i.e.,}$$

$$W''(t) = -\sec^2\left(\frac{\kappa(\alpha)\pi}{2}t\right)\left(\frac{\kappa(\alpha)\pi}{2}\right)^2 + \psi'(\alpha t)\alpha^2 - \csc^2\left(\frac{\pi\alpha}{2}t\right)\left(\frac{\pi\alpha}{2}\right)^2 - \psi'(t) + \csc^2(\pi t)\pi^2 \geq 0.$$

Using series representation of $\psi(x)$ [9, 8.363.8], we can show

$$\psi'(\alpha t)\alpha^2 - \psi'(t) = \sum_{s=0}^{\infty}\frac{\alpha^2}{(\alpha t + s)^2} - \sum_{s=0}^{\infty}\frac{1}{(t+s)^2} = \sum_{s=0}^{\infty}\left(\frac{1}{(t+s/\alpha)^2} - \frac{1}{(t+s)^2}\right) \geq 0,$$

because for now we consider $\alpha > 1$. Thus, it suffices to show that

$$Q(t;\alpha) = -\sec^2\left(\frac{\kappa(\alpha)\pi}{2}t\right)\left(\frac{\kappa(\alpha)\pi}{2}\right)^2 - \csc^2\left(\frac{\pi\alpha}{2}t\right)\left(\frac{\pi\alpha}{2}\right)^2 + \csc^2(\pi t)\pi^2 \geq 0.$$

To show $Q(t;\alpha) \geq 0$, we can treat $Q(t;\alpha)$ as a function of $\alpha$ (for fixed $t$). Because both $\frac{1}{\sin(x)}$ and $\frac{1}{\cos(x)}$ are convex functions of $x \in [0, \pi/2]$, we know $Q(t;\alpha)$ is a concave function of $\alpha$ (for fixed $t$). It is easy to check that

$$\lim_{\alpha \to 1+} Q(t;\alpha) = 0, \qquad \lim_{\alpha \to 2-} Q(t;\alpha) = 0.$$

Because $Q(t;\alpha)$ is concave in $\alpha \in [1, 2]$, we must have $Q(t;\alpha) \geq 0$; and consequently, $W''(t) \geq 0$ and $g'(t) \geq 0$. Therefore, we have proved that (41) decreases monotonically with increasing $k$, when $1 < \alpha \leq 2$.

For $\alpha < 1$ (i.e., $\kappa(\alpha) = \alpha < 1$), we prove the monotonicity by a different technique. First, using the infinite-product representations of Gamma function[9, 8.322] and sin function[9, 1.431.1],

$$\Gamma(z) = \frac{\exp(-\gamma_e z)}{z} \prod_{s=1}^{\infty} \left(1 + \frac{z}{s}\right)^{-1} \exp\left(\frac{z}{s}\right), \qquad \sin(z) = z \prod_{s=1}^{\infty} \left(1 - \frac{z^2}{s^2 \pi^2}\right),$$

we can rewrite (41) as

$$\left[2\cos\left(\frac{\kappa(\alpha)\pi}{2k}\right) \frac{\Gamma\left(\frac{\alpha}{k}\right) \sin\left(\frac{\pi\alpha}{2k}\right)}{\Gamma\left(\frac{1}{k}\right) \sin\left(\frac{\pi}{k}\right)}\right]^k = \left[\frac{\Gamma\left(\frac{\alpha}{k}\right) \sin\left(\frac{\pi\alpha}{k}\right)}{\Gamma\left(\frac{1}{k}\right) \sin\left(\frac{\pi}{k}\right)}\right]^k$$

$$= \exp(-\gamma_e(\alpha-1)) \times \left(\prod_{s=1}^{\infty} \exp\left(\frac{\alpha-1}{sk}\right) \left(1 + \frac{\alpha}{ks}\right)^{-1} \left(1 + \frac{1}{ks}\right) \left(1 - \frac{\alpha^2}{k^2 s^2}\right) \left(1 - \frac{1}{s^2 k^2}\right)^{-1}\right)^k.$$

To show its monotonicity, it suffices to show that for any $s \geq 1$

$$\left(\left(1 + \frac{\alpha}{ks}\right)^{-1} \left(1 + \frac{1}{ks}\right) \left(1 - \frac{\alpha^2}{k^2 s^2}\right) \left(1 - \frac{1}{s^2 k^2}\right)^{-1}\right)^k$$

decreases monotonically, which is equivalent to show the monotonicity of $g(t)$ with increasing $t$, for $t \geq 2$, where

$$g(t) = t \log\left(\left(1 + \frac{\alpha}{t}\right)^{-1} \left(1 + \frac{1}{t}\right) \left(1 - \frac{\alpha^2}{t^2}\right) \left(1 - \frac{1}{t^2}\right)^{-1}\right) = t \log\left(\frac{t-\alpha}{t-1}\right).$$

It is straightforward to show that $t \log\left(\frac{t-\alpha}{t-1}\right)$ is monotonically decreasing with increasing $t$ ($t \geq 2$), for $\alpha < 1$.

To this end, we have proved that for $0 < \alpha \leq 2$ ($\alpha \neq 1$),

$$\left[\cos\left(\frac{\kappa(\alpha)\pi}{2k}\right) \frac{2}{\pi} \Gamma\left(\frac{\alpha}{k}\right) \Gamma\left(1 - \frac{1}{k}\right) \sin\left(\frac{\pi}{2}\frac{\alpha}{k}\right)\right]^k \to \exp(-\gamma_e(\alpha-1)),$$

monotonically with increasing $k$ ($k \geq 2$).

## C  Proof of Lemma 4

We first find the constant $G_{R,gm}$ in the right tail bound

$$\mathbf{Pr}\left(\hat{F}_{(\alpha),gm} - F_{(\alpha)} > \epsilon F_{(\alpha)}\right) \leq \exp\left(-k \frac{\epsilon^2}{G_{R,gm}}\right), \quad \epsilon > 0.$$

For $0 < t < k$, the Markov moment bound yields

$$\mathbf{Pr}\left(\hat{F}_{(\alpha),gm} - F_{(\alpha)} > \epsilon F_{(\alpha)}\right) \leq \frac{\mathbf{E}\left(\hat{F}_{(\alpha),gm}\right)^t}{(1+\epsilon)^t F_{(\alpha)}^t}$$

$$= (1+\epsilon)^{-t} \frac{\left[\cos\left(\frac{\kappa(\alpha)\pi}{2k}t\right) \frac{2}{\pi} \Gamma\left(\frac{\alpha t}{k}\right) \Gamma\left(1 - \frac{t}{k}\right) \sin\left(\frac{\pi\alpha t}{2k}\right)\right]^k}{\left[\cos\left(\frac{\kappa(\alpha)\pi}{2k}\right) \frac{2}{\pi} \Gamma\left(\frac{\alpha}{k}\right) \Gamma\left(1 - \frac{1}{k}\right) \sin\left(\frac{\pi}{2}\frac{\alpha}{k}\right)\right]^{kt}}$$

$$\leq (1+\epsilon)^{-t} \frac{\left[\cos\left(\frac{\kappa(\alpha)\pi}{2k}t\right) \frac{2}{\pi} \Gamma\left(\frac{\alpha t}{k}\right) \Gamma\left(1 - \frac{t}{k}\right) \sin\left(\frac{\pi\alpha t}{2k}\right)\right]^k}{\exp(-t\gamma_e(\alpha-1))}.$$

We need to find the $t$ that minimizes the upper bound. For convenience, we consider its logarithm, i.e.,

$$g(t) = t\gamma_e(\alpha-1) - t\log(1+\epsilon) + k\log\left(\cos\left(\frac{\kappa(\alpha)\pi}{2k}t\right) \frac{2}{\pi} \Gamma\left(\frac{\alpha t}{k}\right) \Gamma\left(1 - \frac{t}{k}\right) \sin\left(\frac{\pi\alpha t}{2k}\right)\right)$$

whose first and second derivatives (with respect to $t$) are

$$g'(t) = \gamma_e(\alpha - 1) - \log(1+\epsilon) - \frac{\kappa(\alpha)\pi}{2}\tan\left(\frac{\kappa(\alpha)\pi}{2k}t\right) + \frac{\alpha\pi/2}{\tan\left(\frac{\alpha\pi t}{2k}\right)} + \psi\left(\frac{\alpha t}{k}\right)\alpha - \psi\left(1 - \frac{t}{k}\right),$$

$$g''(t) = \frac{1}{k}\left(-\left(\frac{\kappa(\alpha)\pi}{2}\right)^2 \sec^2\left(\frac{\kappa(\alpha)\pi}{2k}t\right) - \left(\frac{\alpha\pi}{2}\right)^2 \csc^2\left(\frac{\alpha\pi t}{2k}\right) + \alpha^2\psi'\left(\frac{\alpha t}{k}\right) + \psi'\left(1 - \frac{t}{k}\right)\right),$$

where $\psi(z) = \Gamma'(z)/\Gamma(z)$ is the Psi function.

To show that $g(t)$ is a convex function, i.e., $g''(t) \geq 0$, we make use of the following expansions: [9, 1.422.2, 1.422.4, 8.363.8]

$$\sec^2\left(\frac{\pi x}{2}\right) = \frac{4}{\pi^2}\sum_{j=1}^{\infty}\left(\frac{1}{(2j-1-x)^2} + \frac{1}{(2j-1+x)^2}\right),$$

$$\csc^2(\pi x) = \frac{1}{\pi^2 x^2} + \frac{2}{\pi^2}\sum_{j=1}^{\infty}\frac{x^2 + j^2}{(x^2 - j^2)^2},$$

$$\psi'(x) = \sum_{j=0}^{\infty}\frac{1}{(x+j)^2},$$

to rewrite

$$kg''(t) = -\kappa^2\sum_{j=1}^{\infty}\left(\frac{1}{(2j-1-\kappa t/k)^2} + \frac{1}{(2j-1+\kappa t/k)^2}\right) - \frac{k^2}{t^2} - \frac{\alpha^2}{2}\sum_{j=1}^{\infty}\frac{(\alpha t/2k)^2 + j^2}{((\alpha t/2k)^2 - j^2)^2}$$

$$+ \alpha^2\sum_{j=0}^{\infty}\frac{1}{(\alpha t/k + j)^2} + \sum_{j=0}^{\infty}\frac{1}{(1 - t/k + j)^2}$$

$$= -\kappa^2\sum_{j=1}^{\infty}\left(\frac{1}{(2j-1-\kappa t/k)^2} + \frac{1}{(2j-1+\kappa t/k)^2}\right) - \alpha^2\sum_{j=1}^{\infty}\left(\frac{1}{(\alpha t/k - 2j)^2} + \frac{1}{(\alpha t/k + 2j)^2}\right)$$

$$+ \alpha^2\sum_{j=1}^{\infty}\frac{1}{(\alpha t/k + j)^2} + \sum_{j=1}^{\infty}\frac{1}{(j - t/k)^2}.$$

If $\alpha < 1$, i.e., $\kappa(\alpha) = \alpha$, then

$$kg''(t) = -\alpha^2\sum_{j=1}^{\infty}\left(\frac{1}{(\alpha t/k - j)^2} + \frac{1}{(\alpha t/k + j)^2}\right) + \alpha^2\sum_{j=1}^{\infty}\frac{1}{(\alpha t/k + j)^2} + \sum_{j=1}^{\infty}\frac{1}{(j - t/k)^2}$$

$$= -\alpha^2\sum_{j=1}^{\infty}\frac{1}{(j - \alpha t/k)^2} + \sum_{j=1}^{\infty}\frac{1}{(j - t/k)^2} \geq 0,$$

because $\alpha < 1$ and $0 < t < k$.

If $\alpha > 1$, i.e., $\kappa(\alpha) = 2 - \alpha < 1$, then

$$kg''(t) = -\kappa^2\sum_{j=1}^{\infty}\left(\frac{1}{(2j-1-\kappa t/k)^2} + \frac{1}{(2j-1+\kappa t/k)^2}\right) - \alpha^2\sum_{j=1}^{\infty}\left(\frac{1}{(\alpha t/k - 2j)^2} + \frac{1}{(\alpha t/k + 2j)^2}\right)$$

$$+ \alpha^2\sum_{j=1}^{\infty}\frac{1}{(\alpha t/k + 2j)^2} + \alpha^2\sum_{j=1}^{\infty}\frac{1}{(\alpha t/k + 2j - 1)^2} + \sum_{j=1}^{\infty}\frac{1}{(2j - t/k)^2} + \sum_{j=1}^{\infty}\frac{1}{(2j - 1 - t/k)^2}$$

$$\geq -\kappa^2\sum_{j=1}^{\infty}\frac{1}{(2j-1+\kappa t/k)^2} - \alpha^2\sum_{j=1}^{\infty}\frac{1}{(2j - \alpha t/k)^2} + \alpha^2\sum_{j=1}^{\infty}\frac{1}{(\alpha t/k + 2j - 1)^2} + \sum_{j=1}^{\infty}\frac{1}{(2j - t/k)^2}$$

$$= \left(-\sum_{j=1}^{\infty}\frac{1}{((2j-1)/\kappa + t/k)^2} + \sum_{j=1}^{\infty}\frac{1}{((2j-1)/\alpha + t/k)^2}\right) + \left(-\sum_{j=1}^{\infty}\frac{1}{(2j/\alpha - t/k)^2} + \sum_{j=1}^{\infty}\frac{1}{(2j - t/k)^2}\right)$$

$$\geq 0,$$

because $\alpha > \kappa$.

Since we have proved that $g''(t)$, i.e., $g(t)$ is a convex function, one can find the optimal $t$ by solving $g'(t) = 0$:

$$\gamma_e(\alpha - 1) - \log(1 + \epsilon) - \frac{\kappa(\alpha)\pi}{2}\tan\left(\frac{\kappa(\alpha)\pi}{2k}t\right) + \frac{\alpha\pi/2}{\tan\left(\frac{\alpha\pi t}{2k}\right)} + \psi\left(\frac{\alpha t}{k}\right)\alpha - \psi\left(1 - \frac{t}{k}\right) = 0,$$

We let the solution be $t = C_R k$, where $C_R$ is the solution to

$$\gamma_e(\alpha - 1) - \log(1 + \epsilon) - \frac{\kappa(\alpha)\pi}{2}\tan\left(\frac{\kappa(\alpha)\pi}{2}C_R\right) + \frac{\alpha\pi/2}{\tan\left(\frac{\alpha\pi}{2}C_R\right)} + \psi\left(\alpha C_R\right)\alpha - \psi\left(1 - C_R\right) = 0.$$

Alternatively, we can seek a "sub-optimal" (but asymptotically optimal) solution using the asymptotic expression for $E\left(\hat{F}_{(\alpha),gm}\right)^t$ in Lemma 3:

$$\frac{\left[\cos\left(\frac{\kappa(\alpha)\pi}{2k}t\right)\frac{2}{\pi}\Gamma\left(\frac{\alpha t}{k}\right)\Gamma\left(1 - \frac{t}{k}\right)\sin\left(\frac{\pi\alpha t}{2k}\right)\right]^k}{\left[\cos\left(\frac{\kappa(\alpha)\pi}{2k}\right)\frac{2}{\pi}\Gamma\left(\frac{\alpha}{k}\right)\Gamma\left(1 - \frac{1}{k}\right)\sin\left(\frac{\pi}{2}\frac{\alpha}{k}\right)\right]^{kt}} = \exp\left(\frac{1}{k}\frac{\pi^2}{24}\left(t^2 - t\right)\left(2 + \alpha^2 - 3\kappa^2(\alpha)\right) + ...\right).$$

In other words, we can seek the $t$ that minimizes

$$(1 + \epsilon)^{-t}\exp\left(\frac{1}{k}\frac{\pi^2}{24}\left(t^2 - t\right)\left(2 + \alpha^2 - 3\kappa^2(\alpha)\right)\right),$$

whose minimum is attained at

$$t = k\frac{\log(1 + \epsilon)}{(2 + \alpha^2 - 3\kappa^2(\alpha))\pi^2/12} + \frac{1}{2}.$$

This approximation will produce meaningless bounds even when $\epsilon$ is not too large, especially when $\alpha$ approaches 1. Therefore, despite its simplicity, we do not recommend this sub-optimal constant, which nevertheless can still be quite useful (e.g.,) for serving the initial guess for $C_R$ in a numerical procedure.

Assume we know $C_R$ (e.g., by a simple numerical procedure), we can then express the right tail bound as

$$\mathbf{Pr}\left(\hat{F}_{(\alpha),gm} - F_{(\alpha)} \geq \epsilon F_{(\alpha)}\right) \leq (1 + \epsilon)^{-C_R k}\frac{\left[\cos\left(\frac{\kappa(\alpha)\pi C_R}{2}\right)\frac{2}{\pi}\Gamma\left(\alpha C_R\right)\Gamma\left(1 - C_R\right)\sin\left(\frac{\pi\alpha C_R}{2}\right)\right]^k}{\exp\left(-C_R k\gamma_e(\alpha - 1)\right)}$$

$$= \exp\left(-k\frac{\epsilon^2}{G_{R,gm}}\right),$$

where

$$\frac{\epsilon^2}{G_{R,gm}} = C_R\log(1 + \epsilon) - C_R\gamma_e(\alpha - 1) - \log\left(\cos\left(\frac{\kappa(\alpha)\pi C_R}{2}\right)\frac{2}{\pi}\Gamma\left(\alpha C_R\right)\Gamma\left(1 - C_R\right)\sin\left(\frac{\pi\alpha C_R}{2}\right)\right).$$

Next, we find the constant $G_{L,gm,\alpha,\epsilon,k_0}$ in the left tail bound

$$\mathbf{Pr}\left(\hat{F}_{(\alpha),gm} - F_{(\alpha)} \leq -\epsilon F_{(\alpha)}\right) \leq \exp\left(-k\frac{\epsilon^2}{G_{L,\alpha,\epsilon,k_0}}\right), \quad k > k_0, \quad 0 < \epsilon < 1.$$

From Lemma 3, we know that, for any $t$, where $0 < t < k/\alpha$ if $\alpha > 1$ and $t > 0$ if $\alpha < 1$,

$$\mathbf{Pr}\left(\hat{F}_{(\alpha),gm} \leq (1 - \epsilon)F_{(\alpha)}\right) = \mathbf{Pr}\left(\hat{F}_{(\alpha),gm}^{-t} \geq (1 - \epsilon)^{-t}F_{(\alpha)}^{-t}\right)$$

$$\leq \frac{E\left(\hat{F}_{(\alpha),gm}^{-t}\right)}{(1 - \epsilon)^{-t}F_{(\alpha)}^{-t}} = (1 - \epsilon)^t\frac{\left[-\cos\left(\frac{\kappa(\alpha)\pi}{2k}t\right)\frac{2}{\pi}\Gamma\left(-\frac{\alpha t}{k}\right)\Gamma\left(1 + \frac{t}{k}\right)\sin\left(\frac{\pi\alpha t}{2k}\right)\right]^k}{\left[\cos\left(\frac{\kappa(\alpha)\pi}{2k}\right)\frac{2}{\pi}\Gamma\left(\frac{\alpha}{k}\right)\Gamma\left(1 - \frac{1}{k}\right)\sin\left(\frac{\pi}{2}\frac{\alpha}{k}\right)\right]^{-kt}},$$

which can be minimized (sub-optimally) by finding the $t$, where $t = C_L k$, such that

$$\log(1-\epsilon)C_L - \gamma_e(\alpha-1)C_L + \frac{\kappa(\alpha)\pi}{2}\tan\left(\frac{\kappa(\alpha)\pi}{2}C_L\right) - \frac{\alpha\pi}{2}\tan\left(\frac{\alpha\pi}{2}C_L\right) - \psi(1+\alpha C_L)\alpha + \psi(1+C_L) = 0.$$

Thus, we have shown the left tail bound (for $k > k_0$)

$$\mathbf{Pr}\left(\hat{F}_{(\alpha),gm} - F_{(\alpha)} < -\epsilon F_{(\alpha)}\right) \leq \exp\left(-k\frac{\epsilon^2}{G_{L,gm,k_0}}\right),$$

where

$$\frac{\epsilon^2}{G_{L,gm,k_0}} = -C_L \log(1-\epsilon) - \log\left(-\cos\left(\frac{\kappa(\alpha)\pi}{2}C_L\right)\frac{2}{\pi}\Gamma(-\alpha C_L)\Gamma(1+C_L)\sin\left(\frac{\pi\alpha C_L}{2}\right)\right)$$
$$- k_0 C_L \log\left(\cos\left(\frac{\kappa(\alpha)\pi}{2k_0}\right)\frac{2}{\pi}\Gamma\left(\frac{\alpha}{k_0}\right)\Gamma\left(1-\frac{1}{k_0}\right)\sin\left(\frac{\pi}{2}\frac{\alpha}{k_0}\right)\right).$$

# D  Proof of Lemma 5

From Lemma 4,

$$\frac{\epsilon^2}{G_{R,gm}} = C_R \log(1+\epsilon) - C_R \gamma_e(\alpha-1) - \log\left(\cos\left(\frac{\kappa(\alpha)\pi C_R}{2}\right)\frac{2}{\pi}\Gamma(\alpha C_R)\Gamma(1-C_R)\sin\left(\frac{\pi\alpha C_R}{2}\right)\right),$$

and $C_R$ is the solution to $g_1(C_R, \alpha, \epsilon) = 0$,

$$g_1(C_R, \alpha, \epsilon) = -\gamma_e(\alpha-1) + \log(1+\epsilon) + \frac{\kappa(\alpha)\pi}{2}\tan\left(\frac{\kappa(\alpha)\pi}{2}C_R\right) - \frac{\alpha\pi/2}{\tan\left(\frac{\alpha\pi}{2}C_R\right)} - \psi(\alpha C_R)\alpha + \psi(1-C_R) = 0.$$

Let $\alpha = 1 - \Delta$ if $\alpha < 1$ and $\alpha = 1 + \Delta$ if $\alpha > 1$. Thus, $0 < \Delta < 0$ and $\kappa(\alpha) = 1 - \Delta$.
Using the representations in [9, 1.421.1,1.421.3,8.362.1]

$$\tan\left(\frac{\pi x}{2}\right) = \frac{4x}{\pi}\sum_{j=1}^{\infty}\frac{1}{(2j-1)^2 - x^2},$$

$$\frac{1}{\tan(\pi x)} = \frac{1}{\pi x} + \frac{2x}{\pi}\sum_{j=1}^{\infty}\frac{1}{x^2 - j^2},$$

$$\psi(x) = -\gamma_e - \frac{1}{x} + x\sum_{j=1}^{\infty}\frac{1}{j(x+j)},$$

we rewrite $g_1$ as

$$g_1 = -\gamma_e(\alpha - 1) + \log(1+\epsilon) + \frac{\kappa\pi}{2}\frac{4\kappa C_R}{\pi}\sum_{j=1}^{\infty}\frac{1}{(2j-1)^2 - (\kappa C_R)^2} - \frac{\alpha\pi}{2}\left(\frac{2}{\pi\alpha C_R} + \frac{\alpha C_R}{\pi}\sum_{j=1}^{\infty}\frac{1}{(\alpha C_R/2)^2 - j^2}\right)$$

$$-\alpha\left(-\gamma_e - \frac{1}{\alpha C_R} + \alpha C_R\sum_{j=1}^{\infty}\frac{1}{j(\alpha C_R + j)}\right) + \left(-\gamma_e - \frac{1}{1 - C_R} + (1 - C_R)\sum_{j=1}^{\infty}\frac{1}{j(1 - C_R + j)}\right)$$

$$= \log(1+\epsilon) + 2\kappa^2 C_R\sum_{j=1}^{\infty}\frac{1}{(2j-1)^2 - (\kappa C_R)^2} + 2\alpha^2 C_R\sum_{j=1}^{\infty}\frac{1}{(2j)^2 - (\alpha C_R)^2}$$

$$-\alpha^2 C_R\sum_{j=1}^{\infty}\frac{1}{j(\alpha C_R + j)} + (1 - C_R)\sum_{j=1}^{\infty}\frac{1}{j(1 - C_R + j)} - \frac{1}{1 - C_R}$$

$$= \log(1+\epsilon) + \kappa\sum_{j=1}^{\infty}\left(\frac{1}{2j+1 - \kappa C_R} - \frac{1}{2j-1 + \kappa C_R}\right) + \alpha\sum_{j=1}^{\infty}\left(\frac{1}{2j - \alpha C_R} - \frac{1}{2j + \alpha C_R}\right)$$

$$-\alpha\sum_{j=1}^{\infty}\left(\frac{1}{j} - \frac{1}{\alpha C_R + j}\right) + \sum_{j=1}^{\infty}\left(\frac{1}{j} - \frac{1}{1 - C_R + j}\right) + \frac{\kappa}{1 - \kappa C_R} - \frac{1}{1 - C_R}$$

It is easy to show that, as $\alpha \to 1$, i.e., $\kappa \to 1$, the term

$$\lim_{\alpha \to 1} \kappa\sum_{j=1}^{\infty}\left(\frac{1}{2j+1 - \kappa C_R} - \frac{1}{2j-1 + \kappa C_R}\right) + \alpha\sum_{j=1}^{\infty}\left(\frac{1}{2j - \alpha C_R} - \frac{1}{2j + \alpha C_R}\right)$$

$$-\alpha\sum_{j=1}^{\infty}\left(\frac{1}{j} - \frac{1}{\alpha C_R + j}\right) + \sum_{j=1}^{\infty}\left(\frac{1}{j} - \frac{1}{1 - C_R + j}\right)$$

$$= \lim_{\alpha \to 1} \sum_{j=1}^{\infty}\left(\frac{\kappa}{2j+1 - \kappa C_R} + \frac{\alpha}{2j - \alpha C_R}\right) - \sum_{j=1}^{\infty}\left(\frac{\kappa}{2j-1 + \kappa C_R} + \frac{\alpha}{2j + \alpha C_R}\right)$$

$$-\alpha\sum_{j=1}^{\infty}\left(\frac{1}{j} - \frac{1}{\alpha C_R + j}\right) + \sum_{j=1}^{\infty}\left(\frac{1}{j} - \frac{1}{1 - C_R + j}\right)$$

$$= \lim_{\alpha \to 1} \sum_{j=1}^{\infty}\frac{\kappa}{1 + j - \kappa C_R} - \sum_{j=1}^{\infty}\frac{\kappa}{j + \kappa C_R} - \alpha\sum_{j=1}^{\infty}\left(\frac{1}{j} - \frac{1}{\alpha C_R + j}\right) + \sum_{j=1}^{\infty}\left(\frac{1}{j} - \frac{1}{1 - C_R + j}\right)$$

$$= 0.$$

Recall that, from Lemma 4, we know that $g_1 = 0$ has a unique well-defined solution for $C_R \in (0, 1)$. We also need to analyze the following term

$$\frac{\kappa}{1 - \kappa C_R} - \frac{1}{1 - C_R} = \frac{\kappa - 1}{(1 - \kappa C_R)(1 - C_R)} = \frac{-\Delta}{(1 - \kappa C_R)(1 - C_R)},$$

which, when $\alpha \to 0$, must approach a finite non-zero limit. In other words, We must have $C_R \to 1$, at the rate $O\left(\sqrt{\Delta}\right)$. This argument also provides an approximation for $C_R$ when $\alpha \to 1$, i.e.,

$$C_R = 1 - \sqrt{\frac{\Delta}{\log(1+\epsilon)}} + o\left(\sqrt{\Delta}\right).$$

The next task is to analyze $G_{R,gm}$.

$$\frac{\epsilon^2}{G_{R,gm}} = C_R\log(1+\epsilon) - C_R\gamma_e(\alpha - 1) - \log\left(\cos\left(\frac{\kappa(\alpha)\pi C_R}{2}\right)\frac{2}{\pi}\Gamma(\alpha C_R)\Gamma(1 - C_R)\sin\left(\frac{\pi\alpha C_R}{2}\right)\right)$$

$$= C_R\log(1+\epsilon) - C_R\gamma_e(\alpha - 1) + \log\left(\frac{\cos\left(\frac{\alpha\pi C_R}{2}\right)\Gamma(1 - \alpha C_R)}{\cos\left(\frac{\kappa\pi C_R}{2}\right)\Gamma(1 - C_R)}\right).$$

Using the infinite product representations of the cosine and gamma functions, we can re-write

$$\frac{\cos\left(\frac{\alpha\pi C_R}{2}\right)\Gamma(1-\alpha C_R)}{\cos\left(\frac{\kappa\pi C_R}{2}\right)\Gamma(1-C_R)}$$
$$=\exp(\gamma_e(\alpha-1)C_R)\frac{1-C_R}{1-\alpha C_R}$$
$$\times\prod_{j=0}^{\infty}\left(1-\frac{\alpha^2 C_R^2}{(2j+1)^2}\right)\left(1-\frac{\kappa^2 C_R^2}{(2j+1)^2}\right)^{-1}\prod_{j=1}^{\infty}\exp\left(\frac{(1-\alpha)C_R}{j}\right)\left(1+\frac{1-C_R}{j}\right)\left(1+\frac{1-\alpha C_R}{j}\right)^{-1}$$
$$=\exp(\gamma_e(\alpha-1)C_R)\frac{(1+\alpha C_R)(1-C_R)}{1-\kappa^2 C_R^2}$$
$$\times\prod_{j=1}^{\infty}\left(1-\frac{\alpha^2 C_R^2}{(2j+1)^2}\right)\left(1-\frac{\kappa^2 C_R^2}{(2j+1)^2}\right)^{-1}\exp\left(\frac{(1-\alpha)C_R}{j}\right)\left(1+\frac{1-C_R}{j}\right)\left(1+\frac{1-\alpha C_R}{j}\right)^{-1},$$

Taking logarithm of which yields

$$\log\frac{\cos\left(\frac{\alpha\pi C_R}{2}\right)\Gamma(1-\alpha C_R)}{\cos\left(\frac{\kappa\pi C_R}{2}\right)\Gamma(1-C_R)}$$
$$=\gamma_e(\alpha-1)C_R+\log\frac{(1+\alpha C_R)(1-C_R)}{1-\kappa^2 C_R^2}+\sum_{j=1}^{\infty}\log\frac{\left(1-\frac{\alpha^2 C_R^2}{(2j+1)^2}\right)}{\left(1-\frac{\kappa^2 C_R^2}{(2j+1)^2}\right)}+\left(\frac{(1-\alpha)C_R}{j}\right)+\log\frac{\left(1+\frac{1-C_R}{j}\right)}{\left(1+\frac{1-\alpha C_R}{j}\right)}.$$

If $\alpha < 1$, i.e., $\kappa = \alpha = 1 - \Delta$, then

$$\log\frac{\cos\left(\frac{\alpha\pi C_R}{2}\right)\Gamma(1-\alpha C_R)}{\cos\left(\frac{\kappa\pi C_R}{2}\right)\Gamma(1-C_R)}$$
$$=-\gamma_e\Delta C_R+\log\frac{1-C_R}{1-\alpha C_R}+\sum_{j=1}^{\infty}\left(\frac{(1-\alpha)C_R}{j}\right)+\log\frac{\left(1+\frac{1-C_R}{j}\right)}{\left(1+\frac{1-\alpha C_R}{j}\right)}$$
$$=-\gamma_e\Delta C_R-\log\left(1+\frac{\Delta C_R}{1-C_R}\right)+\sum_{j=1}^{\infty}\left(\frac{1}{2}\left(\frac{1-\alpha C_R}{j}\right)^2-\frac{1}{2}\left(\frac{1-C_R}{j}\right)^2+...\right)$$
$$=-\gamma_e\Delta C_R-\log\left(1+\frac{\Delta C_R}{1-C_R}\right)+\frac{\pi^2}{12}C_R\Delta(2-\alpha C_R-C_R)+...$$

Thus, for $\alpha < 1$, consider $C_R = 1 - \sqrt{\frac{\Delta}{\log(1+\epsilon)}} + o\left(\sqrt{\Delta}\right)$, we have

$$\frac{\epsilon^2}{G_{R,gm}}=C_R\log(1+\epsilon)-\frac{\Delta C_R}{1-C_R}+\frac{\pi^2}{12}C_R\Delta(2-\alpha C_R-C_R)+...$$
$$=\log(1+\epsilon)-2\sqrt{\Delta\log(1+\epsilon)}+o\left(\sqrt{\Delta}\right)$$

If $\alpha > 1$, i.e., $\alpha = 1 + \Delta$ and $\kappa = 1 - \Delta$, then (using above result for $\alpha < 1$)

$$\log\frac{\cos\left(\frac{\alpha\pi C_R}{2}\right)\Gamma(1-\alpha C_R)}{\cos\left(\frac{\kappa\pi C_R}{2}\right)\Gamma(1-C_R)}$$
$$=\gamma_e\Delta C_R+\log\frac{(1+\alpha C_R)(1-C_R)}{1-\kappa^2 C_R^2}+\sum_{j=1}^{\infty}\log\frac{\left(1-\frac{\alpha^2 C_R^2}{(2j+1)^2}\right)}{\left(1-\frac{\kappa^2 C_R^2}{(2j+1)^2}\right)}+...$$

$$\log \frac{(1+\alpha C_R)(1-C_R)}{1-\kappa^2 C_R^2} = \log \frac{1+\alpha C_R}{1+\kappa C_R} - \log \frac{1-\kappa C_R}{1-C_R}$$

$$= \log\left(1 + \frac{2\Delta C_R}{1+\kappa C_R}\right) - \log\left(1 + \frac{\Delta C_R}{1-C_R}\right)$$

$$= -\sqrt{\Delta}\log(1+\epsilon) + o\left(\sqrt{\Delta}\right).$$

$$\sum_{j=1}^{\infty} \log \frac{\left(1-\frac{\alpha^2 C_R^2}{(2j+1)^2}\right)}{\left(1-\frac{\kappa^2 C_R^2}{(2j+1)^2}\right)} = \sum_{j=1}^{\infty} \log \frac{1+\frac{\alpha C_R}{2j+1}}{1+\frac{\kappa C_R}{2j+1}} + \log \frac{1-\frac{\alpha C_R}{2j+1}}{1-\frac{\kappa C_R}{2j+1}}$$

$$= \sum_{j=1}^{\infty} \log\left(1 + \frac{\frac{2\Delta C_R}{2j+1}}{1+\frac{\kappa C_R}{2j+1}}\right) + \log\left(1 - \frac{\frac{2\Delta C_R}{2j+1}}{1-\frac{\kappa C_R}{2j+1}}\right) = O(\Delta).$$

Therefore, for $\alpha > 1$, we also have

$$\frac{\epsilon^2}{G_{R,gm}} = \log(1+\epsilon) - 2\sqrt{\Delta \log(1+\epsilon)} + o\left(\sqrt{\Delta}\right).$$

In other words, as $\alpha \to 1$, the constant $G_{R,gm}$ converges to $\frac{\epsilon^2}{\log(1+\epsilon)}$ at the rate $O\left(\sqrt{\Delta}\right)$, i.e.,

$$G_{R,gm} = \frac{\epsilon^2}{\log(1+\epsilon) - 2\sqrt{\Delta \log(1+\epsilon)} + o\left(\sqrt{\Delta}\right)}.$$

# E  Proof of Lemma 6

Assume $k$ i.i.d. samples $x_j \sim S(\alpha < 1, \beta = 1, F_{(\alpha)})$. Using the $(-\alpha)$th moment in Lemma 1 suggests that

$$\hat{R}_{(\alpha)} = \frac{\frac{1}{k}\sum_{j=1}^{k} |x_j|^{-\alpha}}{\frac{\cos\left(\frac{\alpha\pi}{2}\right)}{\Gamma(1+\alpha)}},$$

is an unbiased estimator of $d_{(\alpha)}^{-1}$, whose variance is

$$\text{Var}\left(\hat{R}_{(\alpha)}\right) = \frac{d_{(\alpha)}^{-2}}{k}\left(\frac{2\Gamma^2(1+\alpha)}{\Gamma(1+2\alpha)} - 1\right).$$

We can then estimate $F_{(\alpha)}$ by $\frac{1}{\hat{R}_{(\alpha)}}$, i.e.,

$$\hat{F}_{(\alpha),hm} = \frac{1}{\hat{R}_{(\alpha)}} = \frac{k\frac{\cos\left(\frac{\alpha\pi}{2}\right)}{\Gamma(1+\alpha)}}{\sum_{j=1}^{k} |x_j|^{-\alpha}}.$$

which is biased at the order $O\left(\frac{1}{k}\right)$. To remove the $O\left(\frac{1}{k}\right)$ term of the bias, we recommend a bias-corrected version obtained by Taylor expansions [15, Theorem 6.1.1]:

$$\frac{1}{\hat{R}_{(\alpha)}} - \frac{\text{Var}\left(\hat{R}_{(\alpha)}\right)}{2}\left(\frac{2}{F_{(\alpha)}^{-3}}\right), \tag{42}$$

from which we obtain the bias-corrected estimator

$$\hat{F}_{(\alpha),hm,c} = \frac{k\frac{\cos\left(\frac{\alpha\pi}{2}\right)}{\Gamma(1+\alpha)}}{\sum_{j=1}^{k} |x_j|^{-\alpha}}\left(1 - \frac{1}{k}\left(\frac{2\Gamma^2(1+\alpha)}{\Gamma(1+2\alpha)} - 1\right)\right), \tag{43}$$

whose bias and variance are

$$\mathrm{E}\left(\hat{F}_{(\alpha),hm,c}\right) = F_{(\alpha)} + O\left(\frac{1}{k^2}\right),$$

$$\mathrm{Var}\left(\hat{F}_{(\alpha),hm,c}\right) = \frac{F_{(\alpha)}^2}{k}\left(\frac{2\Gamma^2(1+\alpha)}{\Gamma(1+2\alpha)} - 1\right) + O\left(\frac{1}{k^2}\right).$$

We now study the tail bounds. For convenience, we provide tail bounds for $\hat{F}_{(\alpha),hm}$ instead of $\hat{F}_{(\alpha),hm,c}$. We first analyze the following moment generating function:

$$\mathrm{E}\left(\exp\left(\frac{F_{(\alpha)}|x_j|^{-\alpha}}{\cos(\alpha\pi/2)/\Gamma(1+\alpha)}t\right)\right)$$

$$=1 + \sum_{m=1}^{\infty} \frac{t^m}{m!} \mathrm{E}\left(F_{(\alpha)}\left(\frac{|x_j|^{-\alpha}}{\cos(\alpha\pi/2)/\Gamma(1+\alpha)}\right)^m\right)$$

$$=1 + \sum_{m=1}^{\infty} \frac{t^m}{m!} \frac{\Gamma(1+m)\Gamma^m(1+\alpha)}{\Gamma(1+m\alpha)}$$

$$=\sum_{m=0}^{\infty} \frac{\Gamma^m(1+\alpha)}{\Gamma(1+m\alpha)} t^m.$$

For the right tail bound,

$$\mathbf{Pr}\left(\hat{F}_{(\alpha),hm} - F_{(\alpha)} \geq \epsilon F_{(\alpha)}\right) = \mathbf{Pr}\left(\frac{k\frac{\cos\left(\frac{\alpha\pi}{2}\right)}{\Gamma(1+\alpha)}}{\sum_{j=1}^{k}|x_j|^{-\alpha}} \geq (1+\epsilon)F_{(\alpha)}\right)$$

$$=\mathbf{Pr}\left(\exp\left(-t\left(\frac{\sum_{j=1}^{k} F_{(\alpha)}|x_j|^{-\alpha}}{\cos(\alpha\pi/2)/\Gamma(1+\alpha)}\right)\right) \geq \exp\left(-t\frac{k}{(1+\epsilon)}\right)\right) \quad (t>0)$$

$$\leq \left(\sum_{m=0}^{\infty} \frac{\Gamma^m(1+\alpha)}{\Gamma(1+m\alpha)}(-t)^m\right)^k \exp\left(t\frac{k}{(1+\epsilon)}\right)$$

$$=\exp\left(-k\left(-\log\left(\sum_{m=0}^{\infty} \frac{\Gamma^m(1+\alpha)}{\Gamma(1+m\alpha)}(-t_1^*)^m\right) - \frac{t_1^*}{1+\epsilon}\right)\right)$$

$$=\exp\left(-k\frac{\epsilon^2}{G_{R,hm}}\right),$$

where $t_1^*$ is the solution to

$$\frac{\sum_{m=1}^{\infty}(-1)^m m(t_1^*)^{m-1}\frac{\Gamma^m(1+\alpha)}{\Gamma(1+m\alpha)}}{\sum_{m=0}^{\infty}(-1)^m (t_1^*)^m \frac{\Gamma^m(1+\alpha)}{\Gamma(1+m\alpha)}} + \frac{1}{1+\epsilon} = 0,$$

which, for numerical reasons, can be written as

$$\sum_{m=1}^{\infty}(-1)^m\left(m(1+\epsilon)\frac{(t_1^*)^{m-1}\Gamma^m(1+\alpha)}{\Gamma(1+m\alpha)} - \frac{(t_1^*)^{m-1}\Gamma^{m-1}(1+\alpha)}{\Gamma(1+(m-1)\alpha)}\right) = 0$$

For the left tail bound,

$$\mathbf{Pr}\left(\hat{F}_{(\alpha),hm} - F_{(\alpha)} \leq -\epsilon F_{(\alpha)}\right) = \mathbf{Pr}\left(\frac{k\frac{\cos\left(\frac{\alpha\pi}{2}\right)}{\Gamma(1+\alpha)}}{\sum_{j=1}^{k}|x_j|^{-\alpha}} \leq (1-\epsilon)F_{(\alpha)}\right)$$

$$=\mathbf{Pr}\left(\exp\left(t\left(\frac{\sum_{j=1}^{k}F_{(\alpha)}|x_j|^{-\alpha}}{\cos(\alpha\pi/2)/\Gamma(1+\alpha)}\right)\right) \geq \exp\left(t\frac{k}{(1-\epsilon)}\right)\right) \qquad (t > 0)$$

$$\leq \left(\sum_{m=0}^{\infty}\frac{\Gamma^m(1+\alpha)}{\Gamma(1+m\alpha)}t^m\right)^k \exp\left(-t\frac{k}{(1-\epsilon)}\right)$$

$$= \exp\left(-k\left(-\log\left(\sum_{m=0}^{\infty}\frac{\Gamma^m(1+\alpha)}{\Gamma(1+m\alpha)}(t_2^*)^m\right) + \frac{t_2^*}{1-\epsilon}\right)\right),$$

where $t_2^*$ is the solution to

$$\sum_{m=1}^{\infty}\left\{\frac{(t_2^*)^{m-1}\Gamma^{m-1}(1+\alpha)}{\Gamma(1+(m-1)\alpha)} - m(1-\epsilon)\frac{(t_2^*)^{m-1}\Gamma^m(1+\alpha)}{\Gamma(1+m\alpha)}\right\} = 0.$$

# F  Proof of Lemma 7

Assume $z \sim S(\alpha = 0.5, \beta = 1, F_{(0.5)})$. For convenience, we will denote $h = F_{(0.5)}$, only in the proof.

The log likelihood, $l(z; h)$, and first three derivatives (w.r.t. $h$) are

$$l(z;h) = \log h - \frac{h^2}{2z} - \frac{3}{2}\log z, \quad l'(z;h) = \frac{1}{h} - \frac{h}{z}, \quad l''(z;h) = -\frac{1}{h^2} - \frac{1}{z}, \quad l'''(z;h) = \frac{2}{h^3}.$$

Therefore, given $k$ i.i.d. samples $x_j \sim S(0.5, 1, h)$, the maximum likelihood estimator (MLE) is computed by

$$\hat{h}_{mle} = \sqrt{\frac{k}{\sum_{j=1}^{k}\frac{1}{x_j}}}.$$

Asymptotically, the variance of the MLE, $\hat{h}_{mle}$ reaches $\frac{1}{kI(h)}$, where $I(h)$ is the Fisher Information:

$$I = I(h) = E(-l''(h)) = \frac{1}{h^2} + E\left(\frac{1}{z}\right).$$

We will soon also need to evaluate higher moments $E\left(\frac{1}{z^m}\right)$. We can utilize the moment generating function of $\frac{1}{z}$, which will be also needed for proving tail bounds in Lemma 8.

$$E\exp\left(\frac{t}{z}\right) = \int_0^{\infty}\frac{h}{\sqrt{2\pi}}\frac{\exp\left(-\frac{h^2}{2z}\right)}{z^{3/2}}\exp\left(\frac{t}{z}\right)dz$$

$$= \frac{h}{\sqrt{2\pi}}\int_0^{\infty}\exp\left(x\left(t - \frac{h^2}{2}\right)\right)x^{-1/2}dx, \qquad \left(x = \frac{1}{z}\right)$$

$$= \frac{h}{\sqrt{2\pi}}\sqrt{\frac{\pi}{h^2/2 - t}} = h\left(h^2 - 2t\right)^{-1/2}, \qquad (t < h^2/2) \qquad [9, 3.472.15]$$

From the $m$th derivative of $E\exp\left(\frac{t}{z}\right)$,

$$\frac{\partial^m E\exp\left(\frac{t}{z}\right)}{\partial t^m} = 1 \times 3 \times 5 \times ... \times (2m-1)h\left(h^2 - 2t\right)^{-\frac{2m+1}{2}}, \qquad m = 1, 2, 3, ...,$$

we can write down

$$\mathrm{E}\left(\frac{1}{z^m}\right) = 1 \times 3 \times 5 \times ... \times (2m-1)h^{-2m}.$$

Therefore, the Fisher Information $\mathrm{I}(h) = \frac{2}{h^2}$. According to the classical statistical results[4, 22], we can obtain the first four moments of $\hat{h}_{mle}$ by evaluating the expressions in [22, 16a-16d],

$$\mathrm{E}\left(\hat{h}_{mle}\right) = d - \frac{[12]}{2k\mathrm{I}^2} + O\left(\frac{1}{k^2}\right)$$

$$\mathrm{Var}\left(\hat{h}_{mle}\right) = \frac{1}{k\mathrm{I}} + \frac{1}{k^2}\left(-\frac{1}{\mathrm{I}} + \frac{[1^4] - [1^22] - [13]}{\mathrm{I}^3} + \frac{3.5[12]^2 - [1^3]^2}{\mathrm{I}^4}\right) + O\left(\frac{1}{k^3}\right)$$

$$\mathrm{E}\left(\hat{h}_{mle} - \mathrm{E}\left(\hat{h}_{mle}\right)\right)^3 = \frac{[1^3] - 3[12]}{k^2\mathrm{I}^3} + O\left(\frac{1}{k^3}\right)$$

$$\mathrm{E}\left(\hat{h}_{mle} - \mathrm{E}\left(\hat{h}_{mle}\right)\right)^4 = \frac{3}{k^2\mathrm{I}^2} + \frac{1}{k^3}\left(-\frac{9}{\mathrm{I}^2} + \frac{7[1^4] - 6[1^22] - 10[13]}{\mathrm{I}^4}\right)$$
$$+ \frac{1}{k^3}\left(\frac{-6[1^3]^2 - 12[1^3][12] + 45[12]^2}{\mathrm{I}^5}\right) + O\left(\frac{1}{k^4}\right),$$

where, after re-formatting,

$$[12] = \mathrm{E}(l')^3 + \mathrm{E}(l'l''), \quad [1^4] = \mathrm{E}(l')^4, \quad [1^22] = \mathrm{E}(l''(l')^2) + \mathrm{E}(l')^4,$$
$$[13] = \mathrm{E}(l')^4 + 3\mathrm{E}(l''(l')^2) + \mathrm{E}(l'l'''), \quad [1^3] = \mathrm{E}(l')^3.$$

Without giving the tails, we report

$$\mathrm{E}\,(l')^3 = -\frac{8}{h^3}, \quad \mathrm{E}\,(l'l'') = \frac{2}{h^3}, \quad \mathrm{E}\,(l')^4 = \frac{60}{h^4}, \quad \mathrm{E}(l''(l')^2) = -\frac{12}{h^4}, \quad \mathrm{E}\,(l'l''') = 0,$$

$$[12] = -\frac{6}{h^3}, \quad [1^4] = \frac{60}{h^4}, \quad [1^22] = \frac{48}{h^4}, \quad [13] = \frac{24}{h^4}, \quad [1^3] = -\frac{8}{h^3}.$$

Thus, we obtain

$$\mathrm{E}\left(\hat{h}_{mle}\right) = h + \frac{3}{4}\frac{h}{k} + O\left(\frac{1}{k^2}\right),$$

$$\mathrm{Var}\left(\hat{h}_{mle}\right) = \frac{1}{2}\frac{h^2}{k} + \frac{15}{8}\frac{h^2}{k^2} + O\left(\frac{1}{k^3}\right),$$

$$\mathrm{E}\left(\hat{h}_{mle} - \mathrm{E}\left(\hat{h}_{mle}\right)\right)^3 = \frac{5}{4}\frac{h^3}{k^2} + O\left(\frac{1}{k^3}\right),$$

$$\mathrm{E}\left(\hat{h}_{mle} - \mathrm{E}\left(\hat{h}_{mle}\right)\right)^4 = \frac{3}{4}\frac{h^4}{k^2} + \frac{93}{8}\frac{h^4}{k^3} + O\left(\frac{1}{k^4}\right).$$

We recommend the bias-corrected version:

$$\hat{h}_{mle,c} = \hat{h}_{mle}\left(1 - \frac{3}{4}\frac{1}{k}\right),$$

whose first four moments, after some algebra, are

$$\mathrm{E}\left(\hat{h}_{mle,c}\right) = h + O\left(\frac{1}{k^2}\right),$$

$$\mathrm{Var}\left(\hat{h}_{mle,c}\right) = \left(1 - \frac{3}{4}\frac{1}{k}\right)^2\left(\frac{1}{2}\frac{h^2}{k} + \frac{15}{8}\frac{h^2}{k^2}\right) + O\left(\frac{1}{k^3}\right) = \frac{1}{2}\frac{h^2}{k} + \frac{9}{8}\frac{h^2}{k^2} + O\left(\frac{1}{k^3}\right),$$

$$\mathrm{E}\left(\hat{h}_{mle,c} - \mathrm{E}\left(\hat{h}_{mle,c}\right)\right)^3 = \frac{5}{4}\frac{h^3}{k^2} + O\left(\frac{1}{k^3}\right),$$

$$\mathrm{E}\left(\hat{h}_{mle,c} - \mathrm{E}\left(\hat{h}_{mle,c}\right)\right)^4 = \left(1 - \frac{3}{4}\frac{1}{k}\right)^4\left(\frac{3}{4}\frac{h^4}{k^2} + \frac{93}{8}\frac{h^4}{k^3}\right) + O\left(\frac{1}{k^4}\right) = \frac{3}{4}\frac{h^4}{k^2} + \frac{75}{8}\frac{h^4}{k^3} + O\left(\frac{1}{k^4}\right).$$

# G  Proof of Lemma 8

Again, for simplicity, we denote only in the proof that $h = F_{(0.5)}$, and hence $\hat{h}_{mle} = \hat{F}_{(0.5),mle}$ etc.

We prove the tail bounds for $\hat{h}_{mle}$, using standard techniques for the Chernoff bounds[5]. For $t > 0$,

$$\mathbf{Pr}\left(\hat{h}_{mle} - h \geq \epsilon h\right) = \mathbf{Pr}\left(\frac{k}{\sum_{j=1}^{k} \frac{1}{x_j}} \geq (1+\epsilon)^2 h^2\right)$$

$$= \mathbf{Pr}\left(-\sum_{j=1}^{k} \frac{1}{x_j} t \geq -t\frac{k-1}{(1+\epsilon)h}\right)$$

$$\leq \left(\prod_{j=1}^{k} \mathbf{E}\left(\exp\left(\frac{-t}{x_j}\right)\right)\right) \exp\left(t\frac{k}{(1+\epsilon)^2 h^2}\right)$$

$$= \left(\frac{h}{(h^2+2t)^{1/2}}\right)^k \exp\left(t\frac{k}{(1+\epsilon)^2 h^2}\right)$$

$$= \exp\left(k \log\left(\frac{h}{(h^2+2t)^{1/2}}\right) + t\frac{k}{(1+\epsilon)^2 h^2}\right),$$

whose minimum is attained at $t = \frac{h^2}{2}\left((1+\epsilon)^2 - 1\right)$. Therefore

$$\mathbf{Pr}\left(\hat{h}_{mle} - h \geq \epsilon h\right) \leq \exp\left(-k\left(\log(1+\epsilon) - \frac{1}{2} + \frac{1}{2}\frac{1}{(1+\epsilon)^2}\right)\right).$$

Similarly, we can prove the left tail bound.

$$\mathbf{Pr}\left(\hat{h}_{mle} - h \leq -\epsilon h\right) = \mathbf{Pr}\left(\frac{k}{\sum_{j=1}^{k} \frac{1}{x_j}} \leq (1-\epsilon)^2 h^2\right)$$

$$= \mathbf{Pr}\left(\sum_{j=1}^{k} \frac{1}{x_j} t \geq t\frac{k}{(1-\epsilon)^2 h^2}\right)$$

$$\leq \left(\prod_{j=1}^{k} \mathbf{E}\left(\exp\left(\frac{t}{x_j}\right)\right)\right) \exp\left(-t\frac{k}{(1-\epsilon)^2 h^2}\right)$$

$$= \left(\frac{h}{(h^2-2t)^{1/2}}\right)^k \exp\left(-t\frac{k}{(1-\epsilon)^2 h^2}\right),$$

whose minimum is attained at $t = \frac{h^2}{2}\left(1 - (1-\epsilon)^2\right)$. Therefore,

$$\mathbf{Pr}\left(\hat{h}_{mle} - h \leq -\epsilon h\right) \leq \exp\left(-k\left(\log(1-\epsilon) - \frac{1}{2} + \frac{1}{2}\frac{1}{(1-\epsilon)^2}\right)\right).$$

For small $\epsilon$, because $\log(1+\epsilon) = \epsilon - \frac{\epsilon^2}{2} + \frac{\epsilon^3}{3}...$ and $\frac{1}{(1+\epsilon)^2} = 1 - 2\epsilon + 3\epsilon^2 + 4\epsilon^3...$, these bounds become

$$\mathbf{Pr}\left(\hat{h}_{mle} - h \geq \epsilon h\right) \leq \exp\left(-k\left(\epsilon^2 - \frac{5}{3}\epsilon^3 + ...\right)\right),$$

$$\mathbf{Pr}\left(\hat{h}_{mle} - h \leq -\epsilon h\right) \leq \exp\left(-k\left(\epsilon^2 + \frac{5}{3}\epsilon^3 + ...\right)\right).$$

# H  Proof of Lemma 9

Assume $k$ i.i.d. samples $x_j \sim S(\alpha, \beta, F_{(\alpha)})$. We first seek an unbiased estimator of $F_{(\alpha)}^\lambda$, denoted by $\hat{R}_{(\alpha),\lambda}$,

$$\hat{R}_{(\alpha),\lambda} = \frac{1}{k} \frac{\sum_{j=1}^k |x_j|^{\lambda\alpha}}{\frac{\cos(\kappa(\alpha)\frac{\lambda\pi}{2})}{|\cos(\frac{\alpha\pi}{2})|^\lambda} \left[\frac{2}{\pi}\Gamma(1-\lambda)\Gamma(\lambda\alpha)\sin\left(\frac{\pi}{2}\lambda\alpha\right)\right]},$$

whose variance is

$$\text{Var}\left(\hat{R}_{(\alpha),\lambda}\right) = \frac{F_{(\alpha)}^{2\lambda}}{k} \left(\frac{\cos(\kappa(\alpha)\lambda\pi)\frac{2}{\pi}\Gamma(1-2\lambda)\Gamma(2\lambda\alpha)\sin(\pi\lambda\alpha)}{\left[\cos(\kappa(\alpha)\frac{\lambda\pi}{2})\frac{2}{\pi}\Gamma(1-\lambda)\Gamma(\lambda\alpha)\sin\left(\frac{\pi}{2}\lambda\alpha\right)\right]^2} - 1\right).$$

In order for the variance to be bounded, we need to restrict $-1/2\alpha < \lambda < 1/2$ if $\alpha > 1$, and $\lambda < 1/2$ if $\alpha < 1$.

A biased estimator of $F_{(\alpha)}$ would be simply $\left(\hat{R}_{(\alpha),\lambda}\right)^{1/\lambda}$, which has $O\left(\frac{1}{k}\right)$ bias. This bias can be removed to an extent by Taylor expansions [15, Theorem 6.1.1].

We call this new estimator the "fractional power" estimator:

$$\hat{F}_{(\alpha),fp,c,\lambda} = \left(\hat{R}_{(\alpha),\lambda}\right)^{1/\lambda} - \frac{\text{Var}\left(\hat{R}_{(\alpha),\lambda}\right)}{2} \frac{1}{\lambda}\left(\frac{1}{\lambda} - 1\right)\left(d_{(\alpha)}^\lambda\right)^{1/\lambda - 2}$$

$$= \left(\frac{1}{k}\frac{\sum_{j=1}^k |x_j|^{\lambda\alpha}}{\frac{\cos(\kappa(\alpha)\frac{\lambda\pi}{2})}{|\cos(\frac{\alpha\pi}{2})|^\lambda}\frac{2}{\pi}\Gamma(1-\lambda)\Gamma(\lambda\alpha)\sin\left(\frac{\pi}{2}\lambda\alpha\right)}\right)^{1/\lambda} \left(1 - \frac{1}{k}\frac{1}{2\lambda}\left(\frac{1}{\lambda} - 1\right)\left(\frac{\cos(\kappa(\alpha)\lambda\pi)\frac{2}{\pi}\Gamma(1-2\lambda)\Gamma(2\lambda\alpha)\sin(\pi\lambda\alpha)}{\left[\cos(\kappa(\alpha)\frac{\lambda\pi}{2})\frac{2}{\pi}\Gamma(1-\lambda)\Gamma(\lambda\alpha)\sin\left(\frac{\pi}{2}\lambda\alpha\right)\right]^2} - 1\right)\right),$$

where we plug in the estimated $F_{(\alpha)}^\lambda$. The asymptotic variance would be

$$\text{Var}\left(\hat{F}_{(\alpha),fp,c,\lambda}\right) = \text{Var}\left(\hat{R}_{(\alpha),c,\lambda}\right)\left(\frac{1}{\lambda}\left(F_{(\alpha)}^\lambda\right)^{1/\lambda - 1}\right)^2 + O\left(\frac{1}{k^2}\right)$$

$$= F_{(\alpha)}^2 \frac{1}{\lambda^2 k}\left(\frac{\cos(\kappa(\alpha)\lambda\pi)\frac{2}{\pi}\Gamma(1-2\lambda)\Gamma(2\lambda\alpha)\sin(\pi\lambda\alpha)}{\left[\cos(\kappa(\alpha)\frac{\lambda\pi}{2})\frac{2}{\pi}\Gamma(1-\lambda)\Gamma(\lambda\alpha)\sin\left(\frac{\pi}{2}\lambda\alpha\right)\right]^2} - 1\right) + O\left(\frac{1}{k^2}\right).$$

The optimal $\lambda$, denoted by $\lambda^*$, is then

$$\lambda^* = \text{argmin}\left\{\frac{1}{\lambda^2}\left(\frac{\cos(\kappa(\alpha)\lambda\pi)\frac{2}{\pi}\Gamma(1-2\lambda)\Gamma(2\lambda\alpha)\sin(\pi\lambda\alpha)}{\left[\cos(\kappa(\alpha)\frac{\lambda\pi}{2})\frac{2}{\pi}\Gamma(1-\lambda)\Gamma(\lambda\alpha)\sin\left(\frac{\pi}{2}\lambda\alpha\right)\right]^2} - 1\right)\right\}.$$

We denote the optimal fractional power estimator $\hat{F}_{(\alpha),fp,c,\lambda^*}$ by $\hat{F}_{(\alpha),op,c}$.

# I  Proof of Lemma 10

We consider only $\alpha < 1$, i.e., $\kappa(\alpha) = \alpha$, To prove that

$$g(\lambda;\alpha) = \frac{1}{\lambda^2}\left(\frac{\cos(\kappa(\alpha)\lambda\pi)\frac{2}{\pi}\Gamma(1-2\lambda)\Gamma(2\lambda\alpha)\sin(\pi\lambda\alpha)}{\left[\cos(\kappa(\alpha)\frac{\lambda\pi}{2})\frac{2}{\pi}\Gamma(1-\lambda)\Gamma(\lambda\alpha)\sin\left(\frac{\pi}{2}\lambda\alpha\right)\right]^2} - 1\right)$$

is a convex function of $\lambda$, where $\lambda < 1/2$, it suffices show that $\frac{\partial^2 g(\lambda;\alpha)}{\partial \lambda^2} > 0$. Here unless we specify $\lambda = 0$, we always assume $\lambda \neq 0$ to avoid triviality. (It is easy to show $\frac{\partial^2 g(\lambda;\alpha)}{\partial \lambda^2} \to 0$ when $\lambda \to 0$.)

Because $\kappa(\alpha) = \alpha$, we simplify $g(\lambda; \alpha)$ (starting with Euler's reflection formula), to be

$$\begin{aligned}
g(\lambda; \alpha) &= \frac{1}{\lambda^2}\left(\frac{\Gamma(1-2\lambda)\Gamma^2(1-\lambda\alpha)}{\Gamma(1-2\lambda\alpha)\Gamma^2(1-\lambda)} - 1\right) \\
&= \frac{1}{\lambda^2}\left(\alpha \frac{\Gamma(-2\lambda)\Gamma^2(-\lambda\alpha)}{\Gamma(-2\lambda\alpha)\Gamma^2(-\lambda)} - 1\right) \\
&= \frac{1}{\lambda^2}\left(\alpha 2^{2\lambda\alpha - 2\lambda} \frac{\Gamma(-\lambda+1/2)\Gamma(-\lambda\alpha)}{\Gamma(-\lambda\alpha+1/2)\Gamma(-\lambda)} - 1\right) \\
&= \frac{1}{\lambda^2}\left(\alpha 2^{2\lambda\alpha - 2\lambda} \prod_{s=0}^{\infty}\left[\left(1 + \frac{1/2}{-\lambda\alpha + s}\right)\left(1 - \frac{1/2}{-\lambda + 1/2 + s}\right)\right] - 1\right) \\
&= \frac{1}{\lambda^2}\left(\alpha 2^{2\lambda\alpha - 2\lambda} \prod_{s=0}^{\infty} \frac{(2s - 2\lambda\alpha + 1)(s - \lambda)}{(s - \lambda\alpha)(2s + 1 - 2\lambda)} - 1\right) \\
&= \frac{1}{\lambda^2}(CM - 1),
\end{aligned}$$

where

$$C = C(\lambda; \alpha) = \alpha 2^{2\lambda\alpha - 2\lambda}, \quad M = M(\lambda; \alpha) = \prod_{s=0}^{\infty} f_s(\lambda; \alpha), \quad f_s(\lambda; \alpha) = \frac{(2s - 2\lambda\alpha + 1)(s - \lambda)}{(s - \lambda\alpha)(2s + 1 - 2\lambda)},$$

and we have used properties of the Gamma function[9, 8.335.1, 8.325.1]:

$$\Gamma(2z) = \frac{2^{2z-1}}{\sqrt{\pi}}\Gamma(z)\Gamma(z + 1/2), \qquad \frac{\Gamma(\alpha)\Gamma(\beta)}{\Gamma(\alpha+\gamma)\Gamma(\beta-\gamma)} = \prod_{s=0}^{\infty}\left[\left(1 + \frac{\gamma}{\alpha + s}\right)\left(1 - \frac{\gamma}{\beta + s}\right)\right].$$

With respect to $\lambda$, the first two derivatives of $g(\lambda; \alpha)$ are (denoting $w = \log(2)(2\alpha - 2)$)

$$\frac{\partial g}{\partial \lambda} = \frac{1}{\lambda^2}\left(-\frac{2}{\lambda}(CM - 1) + \left(w + \sum_{s=0}^{\infty}\frac{\partial \log f_s}{\partial \lambda}\right)CM\right)$$

$$\frac{\partial^2 g}{\partial \lambda^2} = \frac{CM}{\lambda^2}\left(\frac{6}{\lambda^2} + \sum_{s=0}^{\infty}\frac{\partial^2 \log f_s}{\partial \lambda^2} + \left(w + \sum_{s=0}^{\infty}\frac{\partial \log f_s}{\partial \lambda}\right)^2 - \frac{4}{\lambda}\left(w + \sum_{s=0}^{\infty}\frac{\partial \log f_s}{\partial \lambda}\right)\right) - \frac{6}{\lambda^4}.$$

To show $\frac{\partial^2 g}{\partial \lambda^2} > 0$, it suffices to show

$$\frac{\partial^2 g}{\partial \lambda^2}\lambda^4 = 6(CM - 1) + CM\lambda^2\left(\sum_{s=0}^{\infty}\frac{\partial^2 \log f_s}{\partial \lambda^2} + \left(w + \sum_{s=0}^{\infty}\frac{\partial \log f_s}{\partial \lambda}\right)^2 - \frac{4}{\lambda}\left(w + \sum_{s=0}^{\infty}\frac{\partial \log f_s}{\partial \lambda}\right)\right) > 0.$$

Because $(CM)|_{\lambda=0} = 1$ and $(CM)|_{\lambda \neq 0} > 1$ (which is intuitive and will be shown by algebra), it suffices to show

$$T_1(\lambda; \alpha) = 6(CM - 1) + \lambda^2 \sum_{s=0}^{\infty}\frac{\partial^2 \log f_s}{\partial \lambda^2} + \lambda^2\left(w + \sum_{s=0}^{\infty}\frac{\partial \log f_s}{\partial \lambda}\right)^2 - 4\lambda\left(w + \sum_{s=0}^{\infty}\frac{\partial \log f_s}{\partial \lambda}\right) > 0.$$

Because $T_1(\lambda = 0; \alpha) = 0$, it suffices to show $\lambda \frac{\partial T_1}{\partial \lambda} > 0$, where

$$\frac{\partial T_1}{\partial \lambda} = (6CM - 4)\left(w + \sum_{s=0}^{\infty} \frac{\partial \log f_s}{\partial \lambda}\right) - 2\lambda \sum_{s=0}^{\infty} \frac{\partial^2 \log f_s}{\partial \lambda^2} + \lambda^2 \sum_{s=0}^{\infty} \frac{\partial^3 \log f_s}{\partial \lambda^3}$$

$$+ 2\lambda \left(w + \sum_{s=0}^{\infty} \frac{\partial \log f_s}{\partial \lambda}\right)^2 + 2\lambda^2 \left(w + \sum_{s=0}^{\infty} \frac{\partial \log f_s}{\partial \lambda}\right) \sum_{s=0}^{\infty} \frac{\partial^2 \log f_s}{\partial \lambda^2},$$

$$\lambda \frac{\partial T_1}{\partial \lambda} = (6CM - 4)\lambda \left(w + \sum_{s=0}^{\infty} \frac{\partial \log f_s}{\partial \lambda}\right) - 2\lambda^2 \sum_{s=0}^{\infty} \frac{\partial^2 \log f_s}{\partial \lambda^2} + \lambda^3 \sum_{s=0}^{\infty} \frac{\partial^3 \log f_s}{\partial \lambda^3}$$

$$+ 2\lambda^2 \left(w + \sum_{s=0}^{\infty} \frac{\partial \log f_s}{\partial \lambda}\right)^2 + 2\lambda^3 \left(w + \sum_{s=0}^{\infty} \frac{\partial \log f_s}{\partial \lambda}\right) \sum_{s=0}^{\infty} \frac{\partial^2 \log f_s}{\partial \lambda^2}.$$

Because $CM > 1$ and we will soon show $\lambda \left(w + \sum_{s=0}^{\infty} \frac{\partial \log f_s}{\partial \lambda}\right) > 0$, it suffices to show

$$2\lambda \left(w + \sum_{s=0}^{\infty} \frac{\partial \log f_s}{\partial \lambda}\right) - 2\lambda^2 \sum_{s=0}^{\infty} \frac{\partial^2 \log f_s}{\partial \lambda^2} + \lambda^3 \sum_{s=0}^{\infty} \frac{\partial^3 \log f_s}{\partial \lambda^3}$$

$$+ 2\lambda^2 \left(w + \sum_{s=0}^{\infty} \frac{\partial \log f_s}{\partial \lambda}\right)^2 + 2\lambda^3 \left(w + \sum_{s=0}^{\infty} \frac{\partial \log f_s}{\partial \lambda}\right) \sum_{s=0}^{\infty} \frac{\partial^2 \log f_s}{\partial \lambda^2} = \lambda T_2(\lambda; \alpha) > 0,$$

for which it suffices to show $T_2(0; \alpha) = 0$, and

$$\frac{\partial T_2}{\partial \lambda} = \lambda^2 \sum_{s=0}^{\infty} \frac{\partial^4 \log f_s}{\partial \lambda^4} + 2\left(w + \sum_{s=0}^{\infty} \frac{\partial \log f_s}{\partial \lambda}\right)^2 + 8\lambda \left(w + \sum_{s=0}^{\infty} \frac{\partial \log f_s}{\partial \lambda}\right) \sum_{s=0}^{\infty} \frac{\partial^2 \log f_s}{\partial \lambda^2}$$

$$+ 2\lambda^2 \left(\sum_{s=0}^{\infty} \frac{\partial^2 \log f_s}{\partial \lambda^2}\right)^2 + 2\lambda^2 \left(w + \sum_{s=0}^{\infty} \frac{\partial \log f_s}{\partial \lambda}\right) \sum_{s=0}^{\infty} \frac{\partial^3 \log f_s}{\partial \lambda^3} > 0.$$

To this end, we know in order to prove the convexity of $g(\lambda; \alpha)$, it suffices to prove the following:

$$(CM)|_{\lambda=0} = 1, \qquad (CM)|_{\lambda \neq 0} > 1, \qquad \lambda \left(w + \sum_{s=0}^{\infty} \frac{\partial \log f_s}{\partial \lambda}\right) > 0,$$

$$\sum_{s=0}^{\infty} \frac{\partial^2 \log f_s}{\partial \lambda^2} > 0, \qquad \sum_{s=0}^{\infty} \frac{\partial^4 \log f_s}{\partial \lambda^4} > 0, \qquad 4 \sum_{s=0}^{\infty} \frac{\partial^2 \log f_s}{\partial \lambda^2} + \lambda \sum_{s=0}^{\infty} \frac{\partial^3 \log f_s}{\partial \lambda^3} > 0,$$

where

$$\sum_{s=0}^{\infty} \frac{\partial \log f_s}{\partial \lambda} = \sum_{s=0}^{\infty} \left(\frac{-2\alpha}{2s - 2\lambda\alpha + 1} - \frac{1}{s - \lambda} + \frac{\alpha}{s - \lambda\alpha} + \frac{2}{2s + 1 - 2\lambda}\right),$$

$$\sum_{s=0}^{\infty} \frac{\partial^2 \log f_s}{\partial \lambda^2} = \sum_{s=0}^{\infty} \left(\frac{-4\alpha^2}{(2s - 2\lambda\alpha + 1)^2} - \frac{1}{(s - \lambda)^2} + \frac{\alpha^2}{(s - \lambda\alpha)^2} + \frac{4}{(2s + 1 - 2\lambda)^2}\right),$$

$$\sum_{s=0}^{\infty} \frac{\partial^3 \log f_s}{\partial \lambda^3} = \sum_{s=0}^{\infty} \left(\frac{-16\alpha^3}{(2s - 2\lambda\alpha + 1)^3} - \frac{2}{(s - \lambda)^3} + \frac{2\alpha^3}{(s - \lambda\alpha)^3} + \frac{16}{(2s + 1 - 2\lambda)^3}\right),$$

$$\sum_{s=0}^{\infty} \frac{\partial^4 \log f_s}{\partial \lambda^4} = \sum_{s=0}^{\infty} \left(\frac{-96\alpha^4}{(2s - 2\lambda\alpha + 1)^4} - \frac{6}{(s - \lambda)^4} + \frac{6\alpha^4}{(s - \lambda\alpha)^4} + \frac{96}{(2s + 1 - 2\lambda)^4}\right).$$

First, we can show $(CM)|_{\lambda=0} = 1$ and $\left(w + \sum_{s=0}^{\infty} \frac{\partial \log f_s}{\partial \lambda}\right)|_{\lambda=0} = 0$, because

$$CM|_{\lambda=0} = \alpha \lim_{\lambda \to 0} \frac{(1)(-\lambda)}{(-\lambda\alpha)(1)} \prod_{s=1}^{\infty} \frac{(2s + 1)(s)}{(s)(2s + 1)} = 1,$$

and

$$\sum_{s=0}^{\infty} \frac{\partial \log f_s}{\partial \lambda}\bigg|_{\lambda=0} = -2\alpha + 2 + \sum_{s=1}^{\infty}\left(\frac{-2\alpha}{2s+1} - \frac{1}{s} + \frac{\alpha}{s} + \frac{2}{2s+1}\right)$$

$$= -2\alpha + 2 + (\alpha-1)\sum_{s=1}^{\infty}\frac{1}{s(2s+1)} = -(\alpha-1)\log(2) = -w$$

because $\sum_{s=1}^{\infty} \frac{1}{s(2s+1)} = 2 - 2\log(2)$; see [9, 0.234.8]. Therefore, once we have proved $\sum_{s=0}^{\infty} \frac{\partial^2 \log f_s}{\partial \lambda^2} > 0$, $(CM)|_{\lambda\neq 0} > 1$ and $\lambda\left(w + \sum_{s=0}^{\infty} \frac{\partial \log f_s}{\partial \lambda}\right) > 0$ follows immediately.

To show $\sum_{s=0}^{\infty} \frac{\partial^2 \log f_s}{\partial \lambda^2} > 0$, $\sum_{s=0}^{\infty} \frac{\partial^4 \log f_s}{\partial \lambda^4} > 0$, and $4\sum_{s=0}^{\infty} \frac{\partial^2 \log f_s}{\partial \lambda^2} + \lambda \sum_{s=0}^{\infty} \frac{\partial^3 \log f_s}{\partial \lambda^3} > 0$, we make use of Riemanns' Zeta function[9, 9.511,9.521],

$$\zeta(m, q) = \sum_{s=0}^{\infty} \frac{1}{(s+q)^m} = \frac{1}{\Gamma(m)}\int_0^{\infty} \frac{t^{m-1}e^{-qt}}{1 - e^{-t}}dt, \qquad q < 0, \quad m > 1,$$

to rewrite

$$\sum_{s=0}^{\infty} \frac{\partial^2 \log f_s}{\partial \lambda^2} = \sum_{s=0}^{\infty}\left(\frac{-4\alpha^2}{(2s - 2\lambda\alpha + 1)^2} - \frac{1}{(s-\lambda)^2} + \frac{\alpha^2}{(s-\lambda\alpha)^2} + \frac{4}{(2s+1-2\lambda)^2}\right)$$

$$= -\alpha^2 \zeta\left(2, \frac{1}{2} - \lambda\alpha\right) - \frac{1}{\lambda^2} - \zeta(2, 1 - \lambda) + \frac{\alpha^2}{\lambda^2\alpha^2} + \alpha^2\zeta(2, 1 - \lambda\alpha) + \zeta\left(2, \frac{1}{2} - \lambda\right)$$

$$= \int_0^{\infty} \frac{t}{1 - e^{-t}}\left(-\alpha^2 \exp\left(-t\left(1/2 - \lambda\alpha\right)\right) - \exp\left(-t\left(1 - \lambda\right)\right) + \alpha^2 \exp\left(-t\left(1 - \lambda\alpha\right)\right) + \exp\left(-t\left(1/2 - \lambda\right)\right)\right)dt$$

$$= \int_0^{\infty} \frac{t}{1 - e^{-t}}\left(e^{-t/2} - e^{-t}\right)\left(e^{\lambda t} - \alpha^2 e^{\lambda\alpha t}\right) dt$$

$$= \int_0^{\infty} \frac{te^{-t/2}}{1 + e^{-t/2}}\left(e^{\lambda t} - \alpha^2 e^{\lambda\alpha t}\right) dt = \int_0^{\infty} \frac{t}{1 + e^{-t/2}}\left(e^{-t(1/2-\lambda)} - \alpha^2 e^{-t(1/2-\lambda\alpha)}\right) dt$$

Note that $1 \leq 1 + e^{-t/2} \leq 2$ when $t \in [0, \infty)$, and

$$\int_0^{\infty} t\left(e^{-t(1/2-\lambda)} - \alpha^2 e^{-t(1/2-\lambda\alpha)}\right) dt = \frac{1}{(1/2 - \lambda)^2} - \frac{\alpha^2}{(1/2 - \lambda\alpha)^2} = \frac{1}{(1/2 - \lambda)^2} - \frac{1}{(1/2/\alpha - \lambda)^2} > 0$$

because $\lambda < 1/2$, $\alpha < 1$, and $\int_0^{\infty} t^m e^{-pt}dt = m!p^{-m-1}$. This proves that $\sum_{s=0}^{\infty} \frac{\partial^2 \log f_s}{\partial \lambda^2} > 0$.

Similarly,

$$\sum_{s=0}^{\infty} \frac{\partial^4 \log f_s}{\partial \lambda^4} = \sum_{s=0}^{\infty}\left(\frac{-96\alpha^4}{(2s - 2\lambda\alpha + 1)^4} - \frac{6}{(s-\lambda)^4} + \frac{6\alpha^4}{(s-\lambda\alpha)^4} + \frac{96}{(2s+1-2\lambda)^4}\right)$$

$$= -6\alpha^4 \zeta\left(4, \frac{1}{2} - \lambda\alpha\right) - \frac{6}{\lambda^4} - \zeta(4, 1 - \lambda) + \frac{6\alpha^4}{\lambda^4\alpha^4} + 6\alpha^2\zeta(4, 1 - \lambda\alpha) + 6\zeta\left(4, \frac{1}{2} - \lambda\right)$$

$$= \int_0^{\infty} \frac{t^3}{1 + e^{-t/2}}\left(e^{-t(1/2-\lambda)} - \alpha^4 e^{-t(1/2-\lambda\alpha)}\right) dt$$

$$\geq \frac{3!}{2}\left(\frac{1}{(1/2 - \lambda)^4} - \frac{\alpha^4}{(1/2 - \lambda\alpha)^4}\right) > 0.$$

At this point, it is trivial to show $4\sum_{s=0}^{\infty} \frac{\partial^2 \log f_s}{\partial \lambda^2} + \lambda \sum_{s=0}^{\infty} \frac{\partial^3 \log f_s}{\partial \lambda^3} > 0$ if $\lambda > 0$. For $\lambda < 0$, however, we have to use a slightly different approach.

Note that when $\alpha \to 1$, $W = 4\sum_{s=0}^{\infty} \frac{\partial^2 \log f_s}{\partial \lambda^2} + \lambda \sum_{s=0}^{\infty} \frac{\partial^3 \log f_s}{\partial \lambda^3} \to 0$. Therefore, we can treat $W$ as a function of $\lambda$ for fixed $\lambda$. The only thing we need to show is $\frac{\partial W}{\partial \alpha} < 0$ when $\alpha < 1$ and $\lambda < 0$.

$$\begin{aligned}
\frac{\partial W}{\partial \alpha} &= \frac{\partial \left[ 4\sum_{s=0}^{\infty} \frac{\partial^2 \log f_s}{\partial \lambda^2} + \lambda \sum_{s=0}^{\infty} \frac{\partial^3 \log f_s}{\partial \lambda^3} \right]}{\partial \alpha} \\
&= \int_0^{\infty} \frac{e^{-t(1/2-\lambda\alpha)}}{1+e^{-t/2}} \left( 4t \left[ -2\alpha - \alpha^2 \lambda t \right] + \lambda t^2 \left[ -3\alpha^2 - \alpha^3 \lambda t \right] \right) dt \\
&= -\int_0^{\infty} \frac{e^{-t(1/2-\lambda\alpha)}}{1+e^{-t/2}} \left( 8\alpha t + 7\alpha^2 \lambda t^2 + \alpha^3 \lambda^2 t^3 \right) dt \\
&\leq -\frac{1}{2} \int_0^{\infty} e^{-t(1/2-\lambda\alpha)} \left( 8\alpha t + 7\alpha^2 \lambda t^2 + \alpha^3 \lambda^2 t^3 \right) dt \\
&= -\left( \frac{4\alpha}{(1/2-\lambda\alpha)} + \frac{7/2\alpha^2 \lambda}{(1/2-\lambda\alpha)^2} + \frac{\alpha^3 \lambda^2}{(1/2-\lambda\alpha)^3} \right) \\
&= -\frac{\alpha}{(1/2-\lambda\alpha)^3} \left( 4(1/2-\lambda\alpha)^2 + 7/2\alpha\lambda(1/2-\lambda\alpha) + \alpha^2\lambda^2 \right) \\
&= -\frac{\alpha}{(1/2-\lambda\alpha)^3} \left( \left[ \alpha\lambda + \frac{7}{4}\left(\frac{1}{2}-\alpha\lambda\right) \right]^2 + \left( 4 - \left(\frac{7}{4}\right)^2 \right) \left(\frac{1}{2}-\alpha\lambda\right)^2 \right) < 0.
\end{aligned}$$

This completes the proof of the convexity of $g(\lambda; \alpha)$.

Finally, we need to show that $\lambda^* < 0$, where $\lambda^*$ is the solution to $\frac{\partial g}{\partial \lambda} = 0$, or equivalently, the solution to

$$V(\lambda; \alpha) = -2(CM-1) + \lambda \left( w + \sum_{s=0}^{\infty} \frac{\partial \log f_s}{\partial \lambda} \right) CM = 0,$$

provided we discard the trivial solution $\lambda = 0$. Thus, it suffices to show that $V(\lambda; \alpha)$ increases monotonically as $\lambda > 0$, i.e., $\frac{\partial V}{\partial \lambda} > 0$ if $\lambda > 0$. Because

$$\frac{\partial V}{\partial \lambda} = CM \left( 2 \left( w + \sum_{s=0}^{\infty} \frac{\partial \log f_s}{\partial \lambda} \right) + \lambda \sum_{s=0}^{\infty} \frac{\partial^2 \log f_s}{\partial \lambda^2} + \lambda \left( \sum_{s=0}^{\infty} \frac{\partial \log f_s}{\partial \lambda} \right)^2 \right),$$

it suffices to show $\left( w + \sum_{s=0}^{\infty} \frac{\partial \log f_s}{\partial \lambda} \right) > 0$). This is true because we have shown $\lim_{\lambda \to 0} \left( w + \sum_{s=0}^{\infty} \frac{\partial \log f_s}{\partial \lambda} \right) > 0) = 0$ and $\sum_{s=0}^{\infty} \frac{\partial^2 \log f_s}{\partial \lambda^2} > 0$.

This completes the proof that $\lambda^* < 0$ and hence we have completed the proof for this Lemma.